\documentclass[a4paper,11pt]{article}
\usepackage{jheppub}
\usepackage{lineno}
\usepackage{hyperref}
\usepackage{amsmath}
\usepackage{bbm}
\usepackage{bm}
\usepackage{amsfonts}
\usepackage{amssymb}
\usepackage{latexsym}
\usepackage{subfigure}
\usepackage{amsthm}
\usepackage[english]{babel}
\usepackage{float}
\usepackage{slashed}
\usepackage{xcolor} 
\usepackage{verbatim}
\def\bea{\begin{eqnarray}}
\def\eea{\end{eqnarray}}
\def\non{\nonumber}

\arxivnumber{2304.05435}

\date{\today}

\title{\boldmath Investigation of the concurrent effects of ALP-photon and ALP-electron couplings in Collider and Beam Dump Searches }

\author[a,b]{Jia Liu,}
\author[a]{Yan Luo}
\author[b,a]{and Muyuan Song}
\affiliation[a]{School of Physics and State Key Laboratory of Nuclear Physics and Technology, Peking University, \\ Beijing 100871, China}
\affiliation[b]{Center for High Energy Physics, Peking University,\\ Beijing 100871, China}
\emailAdd{jialiu@pku.edu.cn}
\emailAdd{ly23@stu.pku.edu.cn}
\emailAdd{muyuansong@pku.edu.cn}

\abstract{
 Axion-like particles (ALPs) have been studied in numerous experiments to search for their interactions, but most studies have focused on deriving bounds for the single coupling. However, in ultraviolet (UV) models, these couplings can appear simultaneously, and their interplay could have important implications for collider and beam dump searches. In this study, we investigate the concurrent effects of the ALP-photon and ALP-electron couplings in a simplified model and examine how their simultaneous presence modifies existing bounds. We find that modifications to production cross-sections, decaying branching ratios, and the lifetime of the ALP are the major effects. Our results show that low-energy electron-positron colliders such as Belle-II and BaBar are primarily affected by the first two factors, while beam dump experiments such as E137 and NA64 are affected by the cross sections and lifetime. We also consider two UV models - the KSVZ-like model and a lepton-specific version of the DFSZ model - which have only one of the two couplings at tree-level. However, the other coupling can be generated at loops, and our analysis reveals that the simultaneous presence of the two couplings can significantly modify existing bounds on these models for $10^{-3} < m_a < 10$ GeV, especially for beam dump experiments. Overall, our study highlights the importance of considering the concurrent effects of the ALP-photon and ALP-electron couplings in future collider and beam dump analyses.}

\begin{document}
\maketitle
\flushbottom

\section{Introduction}
The axion was proposed to solve the strong CP-violation problem in the Peccei-Quinn theory~\cite{Peccei:1977hh, Peccei:1977ur,Weinberg:1977ma, Wilczek:1977pj, Kim:1979if, Peccei:2006as}. Axion-like particles (ALPs) are expected to be gauge singlet pseudoscalar particles beyond the Standard Model (BSM), which share similar interactions as the axion but with larger parameter space. The QCD axion's mass is generated by the dynamics of the strong force and is fixed by its decay constant. However, when embedding the Standard Model into string theory, it is possible to introduce a variety of axion-like particles (ALPs)~\cite{Arvanitaki:2009fg}, which mass and decay constant can be independent with each other. 
The properties of ALPs, such as their mass and interaction strength with SM particles, can be studied through various experiments. Astronomical/helioscope observations and telescope experiments have put strong constraints in the mass range of ALPs from a few eVs to MeVs such as~\cite{Raffelt:1990yz, Sloan:2016aub, Raffelt:2006cw, Asztalos:2006kz, Kawasaki:2013ae, Dafni:2018tvj} and a summary can be found in~\cite{AxionLimits}. Beamdump experiments have explored the limits of ALP interactions with electrons and photons in the intermediate mass range from MeVs to GeV~\cite{Dobrich:2015jyk, Bjorken:1988as, NA64:2020qwq, Dusaev:2020gxi, NA64:2021aiq}. High-energy collider experiments, like LEP, LHC, Belle-II, and BaBar, have probed the ALP mass range up to several hundreds of GeVs~\cite{Kleban:2005rj, Mimasu:2014nea, Brivio:2017ije, Batell:2009yf, Belle-II:2020jti, BaBar:2014zli}. The limits on ALPs with fermions and gauge bosons are investigated separately in these searches. Previous analyses~\cite{Bauer:2017ris, Bauer:2018uxu, Bauer:2020jbp, Bauer:2021mvw, Arias-Aragon:2022iwl} of ALP searches have typically started with an effective Lagrangian that includes multiple couplings and has explored the ALP couplings for gauge bosons and fermions at the radiative loop level. Although previous studies have examined constraints on individual couplings from eV to GeVs, they have not given sufficient attention to the interplay between these couplings. 

In the search for QCD axions or ALPs within the range where cosmological and stellar constraints are significant, previous studies~\cite{Bauer:2017ris, Bauer:2020jbp, Xiao:2022rxk, Gao:2020wer, DiLuzio:2021qct, Bonilla:2021ufe, Alonso-Alvarez:2018irt, Ertas:2020xcc, Darme:2020sjf, Afik:2023mhj} have explored the correlation between different ALP couplings. These studies have primarily focused on the simultaneous couplings of ALPs to photons and electrons~\cite{ Xiao:2022rxk, Gao:2020wer, Darme:2020sjf}, photons and nucleons~\cite{DiLuzio:2021qct}, as well as different gauge bosons~\cite{Bauer:2017ris, Bonilla:2021ufe, Alonso-Alvarez:2018irt, Ertas:2020xcc, Afik:2023mhj}. Our study, however, focuses specifically on ALP couplings to photons and electrons, similar to the works in refs.~\cite{Xiao:2022rxk, Gao:2020wer}. Nevertheless, there are significant differences between our study and those previous works. Firstly, our study considers ALP masses that are much heavier, and we specifically concentrate on beamdump and collider studies. Although the study presented in ref.~\cite{Ertas:2020xcc} provides valuable insights into the interplay between ALPs and different gauge bosons and discusses the relaxation of astrophysical constraints, it also examines correlations involving different gauge bosons in the context of beamdump experiments. Moreover, it is worth noting that the previous study in ref.~\cite{Darme:2020sjf} focuses on the interplay between ALP-electron and ALP-photon couplings, specifically in the context of invisible axion decay, whereas our research primarily focuses on visible decay channels. 

Furthermore, while the previous study~\cite{Darme:2020sjf} solely analyzes the ALP effective model, our research encompasses both the effective model and its connection with an ultraviolet complete model. In UV complete models, the interactions between ALPs and SM particles are generally correlated, meaning that ALP-fermion and ALP-gauge boson couplings can be generated through nontrivial internal connections such as KSVZ~\cite{Kim:1979if, Shifman:1979if} and DFSZ~\cite{Zhitnitsky:1980tq, Dine:1981rt, Srednicki:1985xd, Sun:2020iim}. Previous studies have extensively explored the couplings of ALPs through renormalization group equations (RGE), as discussed in references such as \cite{Bonilla:2021ufe, Alonso-Alvarez:2018irt, Chala:2020wvs, Gavela:2019wzg,  DiLuzio:2020oah, Giraldo:2020hwl, Song:2023lxf}. These studies have not only examined the impact of ALP couplings on various observables but several have also investigated their correlations with CP-violation and flavor violation, particularly in relation to low-energy observables. In addition, the interplay of multiple couplings plays a crucial role in investigating new physics scenarios, resulting in notable modifications to the constraints compared to single-coupling analyses. For instance, in the context of the ALP solution to the muon anomalous magnetic moment $({g-2})_{\mu}$, the interplay between the muon and photon couplings becomes crucial~\cite{Chang:2000ii, Buen-Abad:2021fwq, Marciano:2016yhf, Cornella:2019uxs, Bauer:2019gfk, Bauer:2021mvw, Liu:2022tqn}. 

In this study, we focus on the concurrence effect of ALP couplings. As a result, we take ALP-electron and ALP-photon couplings as an example and investigate how the concurrence of couplings will affect the existing limits for intermediate ALP mass in electron-positron colliders like Belle-II~\cite{Belle-II:2020jti} and BaBar~\cite{ BaBar:2014zli}, as well as in electron beamdump experiments like E137 and NA64~\cite{Bjorken:1988as, Essig:2010gu, Liu:2017htz, Dusaev:2020gxi}. One immediate consequence is the modification of the cross-section production. 
There are more Feynman diagrams relevant for the cross-section, but we find that the interference effect is less significant than the non-interference terms in the cross-section. Furthermore, the concurrence of couplings has a significant impact on the ALP decay branching ratios and lifetime. As a result, the effects of concurrence of couplings for collider and beamdump experiments cannot be simply inferred from two separate studies in the single coupling scenario. 

Moreover, instead of treating the ALP-electron and ALP-photon couplings as two free parameters, we consider two benchmark ultraviolet (UV) models: the KSVZ-like model, where the ALP-electron coupling arises radiatively from the ALP-photon coupling, and a lepton-specific version of the DFSZ model (DFSZ-like), where the ALP-photon coupling is generated radiatively through the ALP-lepton couplings. These UV models are good examples for single coupling scenario, because they either have ALP-photon coupling or ALP-electron coupling in the tree-level Lagrangian.
Thus, we found that the concurrence of both couplings significantly modify the limits for the KSVZ-like and DFSZ-like models. In electron-positron collider experiments, the DFSZ-like model is more affected than KSVZ-like one, while in the beamdump experiment, the constraints on both models are significantly modified compared to the single coupling scenario.

The structure of the paper is organized as follows: we introduce our simplified ALP model, which focused on ALP-electron and ALP-photon couplings in Section~\ref{sec: model}. We then quantify the effects of the concurrence of the couplings on the cross-section, branching ratios, and decay lifetime. In Section~\ref{sec: UVmodel}, we discuss two benchmark ultraviolet models, the KSVZ-like and lepton-specific DFSZ models, and calculate the effective ALP-photon or ALP-electron couplings in these models. In Section~\ref{sec: experiment results}, we focus on searches for ALPs with masses in the range $10^{-3} < m_a < 10$ GeV at low-energy electron-positron colliders (Belle-II and BaBar) and electron beamdump experiments (E137 and NA64). We reinterpret the existing constraints in the context of the concurrence scenario and discuss how the two UV models are affected. Finally, in Section~\ref{sec: conclu}, we conclude.

\section{Low energy ALP model with photon and electron couplings  }
\label{sec: model}

To begin with, we start with a simplified ALP (denoted as $a$) effective Lagrangian, interacting with the SM photon and derivative couplings with electrons, for couplings $g^{\text{eff}}_{a\gamma\gamma}$ and $g^{\text{eff}}_{a\bar{e}e}$ respectively,
\bea \label{eq:simpleALPlag}
\mathcal{L}_{a} &&= \frac{1}{2} \partial_\mu a \partial^{\mu} a - \frac{1}{2}m^2_a a^2 - \frac{1}{4}g^{\text{eff}}_{a\gamma\gamma}  \times a  F_{\mu\nu} \tilde{F}^{\mu\nu} + \frac{1}{2} g^{\text{eff}}_{a\bar{e}e} \times \partial_{\mu} a\, \bar{e} \gamma^{\mu} \gamma_5 e,
\eea
where the superscript of couplings (e.g., $g^\text{eff}_{a\gamma\gamma}$) means the coupling includes tree-level interactions and 1-loop radiative corrections from other interactions. We will later use the couplings superscript “0” (e.g., $g^0_{a\gamma\gamma}$ ) to denote the tree-level couplings from the ultraviolet model. $F_{\mu\nu}$ denotes the field tensor of electromagnetic gauge boson and its dual $\tilde{F}^{\mu\nu}$ will be $\frac{1}{2}\epsilon^{\mu\nu\alpha\beta} F_{\alpha\beta} $. The decay width of ALP to electron and positron can be expressed as follows:
\bea\label{eq:axiontoee}
\Gamma_{a \to e \bar{e}} &=&
 \frac{(g^{\text{eff}}_{a\bar{e}e})^2 m^2_e m_a }{8 \pi} \left(1- \frac{4m^2_e}{m^2_a}\right)^{\frac{1}{2}}
 ,
\eea
where the kinematic threshold $m_a > 2 m_e$ should be satisfied. The decay of the ALP into fermion pairs is proportional to the square of the fermion masses, and the expression for other lepton channels, such as $\mu$ and $\tau$, are similar to eq.~(\ref{eq:axiontoee}). 
Moreover, the ALP can decay into a pair of photons and the decay width can be expressed as follows:
\bea \label{eq:axiontoAA}
\Gamma_{a \to \gamma \gamma} &=& 
 \frac{ (g^{\text{eff}}_{a\gamma\gamma} )^2 m^3_a}{64 \pi},
\eea
which is the only decay channel for $m_a < 2 m_e$. When $m_a > 2m_e$, both decay channels are kinematically allowed, thus the total decay width increases which reduces its lifetime. 
The ratio of the decay rates, $\Gamma(a\to \gamma\gamma)/\Gamma(a\to \bar{e}e)$, will depend on the square of the effective couplings and the ALP mass. 
In this study, we focus on the axion mass range $1~{\rm MeV}< m_a < 10~{\rm GeV}$, thus for most of the time we have $m_a \gg 2m_e$. Therefore, the two decay branching ratios satisfy the relation 
\bea
	\frac{{\rm BR}(a \to \gamma \gamma)}{{\rm BR}(a \to e \bar{e})} \approx  \frac{(g^{\text{eff}}_{a\gamma\gamma})^2 m^2_a }{8 (g^{\text{eff}}_{a\bar{e}e})^2  m^2_e}.
\eea
At low-energy electron-positron colliders, ALPs with masses around $\mathcal{O}(1)$ GeV - much larger than $m_e$ - will typically decay into diphotons when the two effective couplings are of comparable magnitude. However, for beam dump experiments with ALP masses around $\mathcal{O}(10)$ MeV, ALP can dominantly decay into an electron-positron pair in some regions of parameter space.
In this section, we have finished the introduction of a simplified ALP model with two couplings at low energy in Effective Field Theory description. Next, we will introduce two ultraviolet models and find out the relations between effective couplings and the UV parameters.

\section{Specific UV models for ALP with electron-ALP and photon-ALP couplings }\label{sec: UVmodel}
In this section, we present two candidate models that illustrate the features of effective ALP couplings in ultraviolet complete models. In general, the UV models can have enough degree of freedom to make the two couplings, ALP-photon and ALP-electron couplings, fully independent parameters. In this case, the analysis at low energy follows the simplified ALP EFT model. 
However, coupling between ALP and photon as well as electron can be correlated, even if the two exhibit independent features at high energies.
Therefore, we focus on ultraviolet models which has only one coupling, either ALP-photon or ALP-electron couplings, while the other coupling is missing at high energy but will be generated radiatively at low energy.  We will demonstrate that the generated coupling can have significant impact for constraints on certain parameter spaces. 
In a general effective ALP model, the ALP Lagrangian can be expressed as follows~\cite{DiLuzio:2020wdo}:
\bea\label{eq:ALP effective Lagrangian}
 \mathcal{L}_{a} \supset N \frac{\alpha_s }{4 \pi} \frac{ a}{v_a} G \tilde{G} + E \frac{\alpha_{\text{QED}} }{4 \pi} \frac{ a}{v_a} F \tilde{F} + \frac{\partial_\mu a}{v_a}J^{\mu}_{\text{PQ}},
 \eea
where the  $U(1)_{\text{PQ}}$ symmetry breaking happens with the VEV $v_a$. The first two terms in the equation involve the electromagnetic and color anomaly coefficients, $E$ and $N$, which can have different values in different models. The SM fermion chiral current, $J^{\text{PQ}}_{\mu}$, can be represented as $J^{\text{PQ}}_{\mu} |_{f_L} = - \overline{f}_L \mathcal{\chi}_{f_L} \gamma^{\mu} f_L$ for chiral fermion $f_L$ as an example~\cite{DiLuzio:2020wdo}, where $ \mathcal{\chi}_{f_L}$ is the PQ charge. The vacuum expectation value $v_a$ can be expressed as $v_a = 2 f_a N$, normalizing the $G\tilde{G}$ term in eq.~(\ref{eq:ALP effective Lagrangian}) and rewritten the equation as~\cite{DiLuzio:2020wdo}:
\bea
 \mathcal{L}_{a}& \supset&  \frac{\alpha_s }{8 \pi}\frac{a}{f_a} G \tilde{G} + \frac{\alpha_{\text{QED}} E}{8 \pi N} \frac{ a}{f_a} F \tilde{F} - \frac{\partial_{\mu} a }{2 f_a N} (\overline{f}_L \mathcal{\chi}_{f_L} \gamma^{\mu} f_L + \overline{f}_R \mathcal{\chi}_{f_R} \gamma^{\mu} f_R) \nonumber \\ 
 &=& \frac{a}{f_a} \frac{\alpha_s}{8\pi} G\tilde{G} + \frac{1}{4} \frac{\alpha_{\text{QED}}E}{2\pi N} \frac{a}{f_a} F\tilde{F} + \frac{\partial_{\mu} a}{2 f_a } c^0_{a\bar{f}f} \overline{f} \gamma^{\mu} \gamma_5 f,
\eea
Thus, the tree-level axion-photon coupling could be read out:
\bea\label{eq:treeALPtophoton}
g^{0}_{a\gamma\gamma} = \frac{\alpha_{\text{QED}}}{2\pi f_a} \frac{E}{N}.
\eea
While tree-level axion-fermion coupling will be:
\bea\label{eq:treeALPtoelectron}
g^0_{a\bar{f}f} =\frac{c^0_{a\bar{f}f}}{f_a}= \frac{\chi_{f_L} -\chi_{f_R}}{2Nf_a} = \frac{\chi_{H_f}}{2Nf_a},
\eea
Where term $\chi_{H_f}$ is from the relevant Yukawa term and carries the PQ charges based on $\chi_{f_L}$ (left-handed) and $\chi_{f_R}$ (right-handed) components. Later, we will introduce some ultraviolet ALP models with photons or SM fermions couplings generated not at the tree level but through radiative processes. In the next section, we will focus on the KSVZ-like model and the lepton specific DFSZ model, which starts with ALP-photon and ALP-electron and generates the other coupling.

\subsection{KSVZ-like model}

\begin{figure}[t!]
\hspace{1cm}
 \includegraphics[scale=.75]{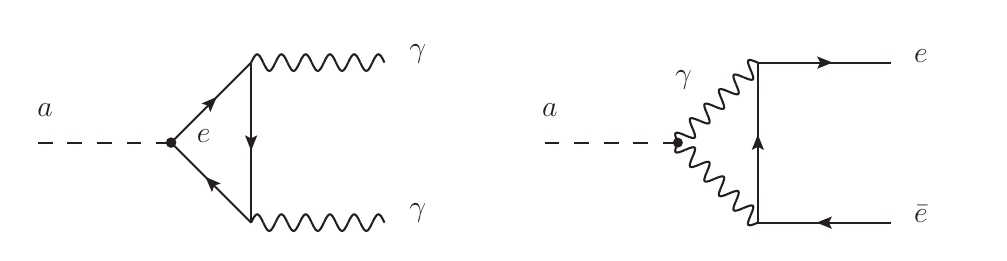}
    \caption{The one-loop Feynman diagram of the contributions to $g_{a\gamma\gamma}^{\text{1-loop}}$ (left) and $g_{a\bar{e}e}^{\text{1-loop}}$ (right). An additional diagram for $g_{a\gamma\gamma}^{\text{1-loop}}$ with exchanged momentum of final states is not shown. 
    } \label{fig:axionloopeepair}
\end{figure}

The origin KSVZ model has already been known for several decades~\cite{Kim:1979if, Shifman:1979if}. To solve the strong CP problem, the minimal KSVZ model adds one vector-like fermion ($\mathcal{Q} =\mathcal{Q}_L + \mathcal{Q}_R$) which is charged under $SU(3)_C$. For our purpose, the vector-like fermion carry SM gauge charge $\sim (1,1,Y)$ for $SU(3)_C$, $SU(2)_L$ and $U(1)_Y$ respectively. In addition, one adds a new singlet complex scalar field $\Phi \sim (1,1,0)$ into the model. The corresponding Lagrangian for $\mathcal{Q}_L$, $\mathcal{Q}_R$ and $\Phi$ satisfies the global $U(1)_{\text{PQ}}$ symmetry, and their PQ charges are given as $+1$, $-1$ and $+2$ in this study, respectively. The Lagrangian and the potential for scalar field can be written as~\cite{DiLuzio:2020wdo}:
\bea
	\mathcal{L}\supset |\partial^\mu \Phi |^2 + i\bar{\mathcal{Q}}\slashed{D}\mathcal{Q} - (\mathcal{Y}_\mathcal{Q} \bar{\mathcal{Q}}_L \mathcal{Q}_R \Phi + {\rm h.c}) - V(\Phi), \,\,  \label{eq:ksvz_Lag} 
 \eea
 \bea
	\,\,\,\,\,\, V(\Phi)  = \lambda_{\Phi} \left(\Phi^* \Phi  - \frac{v^2_a}{2}\right)^2 .
\eea
The global symmetry is spontaneous symmetry broken with VEV $v_a$ and the  complex scalar $\Phi$ can be written as follows:
\bea
\Phi = \frac{1}{\sqrt{2} } (v_a + \rho_a) e^{i\frac{a}{v_a}},
\eea
where the radial component $\rho_a$ obtains a mass of $\sqrt{2 \lambda_\Phi }v_a$, and the singlet pseudoscalar $a$ plays the role of the Goldstone mode and becomes the axion after the PQ symmetry breaking. In the presence of the symmetry breaking, the vector-like fermion $\mathcal{Q}$ in the Lagrangian eq.~(\ref{eq:ksvz_Lag}) acquire masses through its Yukawa interactions, with values around $m_\mathcal{Q} \sim \frac{\mathcal{Y}_\mathcal{Q} v_a}{\sqrt{2}}$. 
The Yukawa term has an exponential term:
\bea
\mathcal{L}\supset -m_\mathcal{Q}\bar{\mathcal{Q}}_L \mathcal{Q}_R e^{i\frac{a}{v_a}}+{\rm h.c.},
\eea
and one can perform the chiral rotation ($\mathcal{Q} \to e^{-i\gamma^5 \frac{a}{2 v_a}} \mathcal{Q}$) as follows,
\bea
\label{eq: axionfieldtransform}
\mathcal{Q}_L \to e^{i \frac{a}{2 v_a}} \mathcal{Q}_L,\,\, \mathcal{Q}_R \to e^{-i \frac{a}{2 v_a}} \mathcal{Q}_R\,\, ,
\eea
to remove the ALP field in the Yukawa Lagrangian. But due to the chiral anomaly, one obtains the following anomalous Lagrangian after electroweak symmetry breaking 
\bea
\delta\mathcal{L}_{\text{KSVZ-like}}&=&\displaystyle{}
Y^2\frac{e^2}{16\pi^2}\frac{a}{v_a}F\tilde{F}\nonumber\\
&=&  E\frac{\alpha_{\text{QED}}}{4 \pi} \frac{a}{v_a} F \tilde{F} ,
\eea
where we use conventional anomaly coefficient $E=Y^2$ for electromagnetic anomaly while the vector fermions do not transform under $SU(3)_C$ in our set up, and we label it as KSVZ-like model.
Therefore, in the KSVZ-like model, we obtain the direct effective ALP-photon coupling from electromagnetic anomaly only: 
\bea
g^{\text{eff}}_{a\gamma\gamma} = E\frac{\alpha_{\text{QED}}}{ \pi} \frac{1}{v_a}.
\eea
Moreover, the ALP-electron coupling could be generated radiatively through the ALP-photon interaction, as shown in the right panel of figure~\ref{fig:axionloopeepair}. 
 The generated ALP-electron coupling is given as~\cite{DiLuzio:2020wdo, Georgi:1986df, Srednicki:1985xd,Chang:1993gm,Bauer:2017ris}:
\bea\label{eq:KSVZinducedaeefromaAA}
g^{\text{eff}}_{a\bar{e}e} = g^{0}_{a\bar{e}e}+ \frac{3\alpha_{\text{QED}}}{4\pi} g^{\text{eff}}_{a\gamma\gamma} \bigg[  \ln \left( \frac{f_a^2}{m_e^2} \right)  + g\left(\tau_e \right)  \bigg], 
\eea
where the value of $f_a/m_e$ is obtained by running the renormalization group from the PQ scale to the scale of the electron mass, with $ g^{0}_{a\bar{e}e}= 0$ for no tree-level contribution. $g(\tau_e)$ is the finite loop term with photon coupling as the right handed of figure~\ref{fig:axionloopeepair}. $g(\tau_e)$ will expressed as follows~\cite{Bauer:2017ris}:
\bea
g(\tau_e) = -\frac{1}{6} \left( \ln \left(\frac{m^2_a}{m^2_{e}} \right) - i\pi \right)^2+\frac{2}{3} +\mathcal{O} \left(\frac{m^2_e}{m^2_a}\right) , \,m^2_a \gg m^2_{e} ,
\eea
when $m^2_a \gg m^2_{e}$, $\mathcal{O} \left(\frac{m^2_e}{m^2_a}\right) $ approaches to zero. 
Contributions to the effective coupling $g^{\text{eff}}_{a\bar{e}e}$ can also come from $F\tilde{Z}$ and $Z\tilde{Z}$, but their effects are only significant for larger values of $f_a$ and $m_a$, which fall outside the scope of this study. Furthermore, as shown in figure~\ref{fig:UVratiowithMa}, any modifications due to these contributions are limited to a maximum of approximately 25\%. As the dominant contribution to $g^{\text{eff}}_{a\bar{e}e}$ comes from $g^{\text{eff}}_{a\gamma\gamma}$, we will disregard these other contributions in our subsequent discussion to maintain a simple relation between the two couplings.

Therefore, we have the ratio between the effective ALP-photon and the effective ALP-electron coupling follows the relation without factor $E$,
\bea\label{eq:ksvzcaAAovercaee}
\frac{|g^{\text{eff}}_{a\gamma\gamma}|}{|g^{\text{eff}}_{a\bar{e}e}|} = \frac{8 \pi}{ \alpha_{\text{QED}} \left| \left(6\ln \left(\frac{f^2_a}{m^2_e}\right)- \left(\ln \left(\frac{m_a^2}{m_e^2}\right)-i
   \pi \right)^2+4\right)\right|} ,
\eea
where $|g^{\text{eff}}_{a\bar{e}e}|$ is naturally smaller due to the loop suppression.
In this model, the new physics degree of freedom at low energy is ALP only, because the radial mode of $\Phi$ and vector-like fermion $\mathcal{Q}$ are quite heavy. Therefore, one can focus on the ALP phenomenology and neglect the possible effects from other particles.

\subsection{DFSZ-like models }

Another intriguing ultraviolet model for the invisible QCD axion is the well-known DFSZ model~\cite{Zhitnitsky:1980tq, Dine:1981rt, Sun:2020iim}, where a new Higgs doublet is involved. Unlike usual DFSZ model, our model based on the lepton-specific setup of 2HDM or termed type-X 2HDM~\cite{Grossman:1994jb, Akeroyd:1994ga, Akeroyd:1996he, Aoki:2009ha, Branco:2011iw, Bhattacharyya:2015nca, Liu:2018xkx}. The model consists of two scalar fields, $H_u \sim (1,2,1/2)$ and $H_d \sim (1,2,1/2)$, which are SU$(3)_c$ singlets and SU$(2)_L$ doublets, respectively. 
In addition, the model contains a SM gauge singlet complex scalar $\Phi$, similar as KSVZ model. We write them as:
\bea
H_{u,d}=\left(
	\begin{array}{c}
		\phi_i^+ \\
		\frac{1}{\sqrt{2}}(v_i+\phi_i+ia_i)
	\end{array}
	\right)
	,(i=1,2)
 \eea
 \bea
 S= \frac{1}{\sqrt{2}}(v_s +\phi_S + i a_s),
\eea 
Moreover, we assign PQ charges to these scalar fields that $X_1=-1$, $X_2=+1$, and $X_s=-X_1+X_2=+2$, respectively. There are two pseudoscalar Goldstone bosons corresponding to broken $SU(2)_Y \times U(1)_Y \times U(1)_{PQ} \to U(1)_{EM}$. In the interacting-eigenstates basis $(a_1,a_2,a_s)$, the Goldstone $G_z$ eaten by vector gauge boson $Z$ is $(v_1,v_2,0)$, and Goldstone $G_A$ corresponding to breaking $U(1)_{PQ}$ is $(X_1v_1,X_2v_2,X_sv_s)$. The physical axion $a$ will be a linear combination between $G_z$ and $G_A$  with $a=c_1 G_z+ c_2 G_A$ and $a\cdot G_z =0$~\cite{Sun:2020iim}:
	\bea
		a \sim (-(X_2-X_1)v^2_2v_1,(X_2-X_1)v^2_1v_2,X_sv_{\text{SM}}^2v_s) \label{dfsz_a},
	\eea
where $v_{\text{SM}}=\sqrt{v_1^2+v_2^2}$. So we get the linear combination of the physical states
are 
\bea
\left(
\begin{array}{c}
  G_z     \\
  a      \\
  A_0 
\end{array}
\right) = 
\left(
\begin{array}{ccc}
 \frac{v_1}{v_{\text{SM}}}     & \frac{v_2}{v_{\text{SM}}} & 0\\
   \frac{v_1 v^2_2}{v_{\text{SM}}\sqrt{v^2_1v^2_2 +v^2_{\text{SM}} v^2_s}}  &  -\frac{v^2_1 v_2}{v_{\text{SM}}\sqrt{v^2_1v^2_2 +v^2_{\text{SM}} v^2_s}} & - \frac{v_s v^2_{\text{SM}}}{v_{\text{SM}}\sqrt{v^2_1v^2_2 +v^2_{\text{SM}} v^2_s}} \\
    \frac{v_2 v_s}{\sqrt{v^2_1v^2_2+v^2_{\text{SM}} v^2_s}} & - \frac{v_1 v_s}{\sqrt{v^2_1v^2_2+v^2_{\text{SM}} v^2_s}} &  \frac{v_1 v_2}{\sqrt{v^2_1v^2_2+v^2_{\text{SM}} v^2_s}}
\end{array}
\right)
\left(
\begin{array}{c}
     a_1  \\
     a_2 \\
     a_s
\end{array}
\right) ,
\eea
where $A_0$ is the CP-odd heavy Higgs in the 2HDM.

We assign PQ charge to quarks of $X_{u_R}=-1$, $X_{d_R}=+1$ and $X_{e_R}=-1$, so the Yukawa interactions are
	\bea
		\mathcal{L}_Y &\supset& -\overline{Q_L}Y_u \tilde{H}_u u_R -\overline{Q_L}Y_d H_u d_R-\overline{L_L}Y_{\ell} H_d e_R +{\rm h.c.},
	\eea
that in our lepton specific setup $H_d$ only couples to leptons and $\tilde{H}_u = i \sigma_2 {H_u}^{*}$. We will take the SM lepton to be $e$, $e,\mu$ and $e,\mu,\tau$ for three different cases which are denoted as DFSZ-like $(e)$, $(e,\mu)$ and $(e,\mu,\tau)$ respectively. Quarks couple only to $H_u$ to obtain the mass. We can get the axion-fermion interaction as 
	\bea \label{eq:dfszgamma5af}
		\mathcal{L}_f\supset& - &i \left(\frac{1}{3}\cos^2\beta\right)\frac{a}{f_a} \bar{u}m_u\gamma_5 u  +i \left(\frac{1}{3}\cos^2\beta\right) \frac{a}{f_a} \bar{d}m_d\gamma_5 d \nonumber \\
		&-&i \left(\frac{1}{3}\sin^2\beta\right) \frac{a}{f_a} \bar{\ell}m_{\ell}\gamma_5 \ell,
	\eea
	where $m_{u}=\frac{1}{\sqrt{2}}Y_u v_1$, $m_{d}=\frac{1}{\sqrt{2}}Y_d v_1$, $m_{e}=\frac{1}{\sqrt{2}}Y_{e} v_2$ 
	and $\tan\beta \equiv v_1/v_2$. The broken scale $f_a$ is defined by
	\bea
		\frac{1}{f_a} \equiv \frac{N_g v_{\text{SM}}}{\sqrt{v_1^2 v_2^2+v_{\text{SM}}^2 v_s^2}},
	\eea
where $N_g=3$ is the generation number. We note the ratio between the coupling with quarks and the coupling with leptons is $1/\tan^2\beta$, thus one can tune the coupling strength between quarks and leptons via $\tan\beta$. After performing the chiral rotation, the $U(1)_{\text{EM}}$ axial anomaly leads to the appearance of the $F\tilde{F}$ term. The $\gamma_5$ term $im_f \bar{f} \gamma_5 f$ in eq.~(\ref{eq:dfszgamma5af}) can then be expressed as follows~\cite{Bauer:2017ris, Bauer:2021mvw}:
	\bea\label{eq:dfszlikegammamugamma5FFtilde}
		\mathcal{L} &\supset& \frac{c_u}{2} \frac{\partial_\mu a}{f_a} \bar{u}\gamma^\mu\gamma_5 u - \frac{c_d}{2} \frac{\partial_\mu a}{f_a} \bar{d}\gamma^\mu\gamma_5 d +	\frac{c_\ell}{2}\frac{\partial_\mu a}{f_a} \bar{\ell}\gamma^\mu\gamma_5 \ell \non\\
		&-&N_g(N^q_c Q^2_{u} c_{u} + N^q_c Q^2_{d} c_{d}+ N^e_c Q^2_\ell c_\ell)\frac{e^2}{16\pi^2}\frac{a}{f_a}F_{\mu\nu}\tilde{F}^{\mu\nu},
	\eea
where $c_{u}=-c_{d}=\frac{1}{3}\cos^2\beta$ and $c_{\ell}=\frac{1}{3}\sin^2\beta$.

$N^q_c$ is the number of colour (3 for quarks and 1 for leptons), $N_g$ is the number of generation and $Q_{i}, i= u,d,\ell$ are the electric charges for quarks and leptons respectively. The anomaly coefficient ($E$) will be the second term of eq.~\ref{eq:dfszlikegammamugamma5FFtilde}, that $E = N_g (N^q_c Q^2_{u} c_{u} + N^q_c Q^2_{d} c_{d}+ N^e_c Q^2_\ell c_\ell) = 1$ in the lepton-specific DFSZ-like model. 
On the other hand,
the axial anomaly for SU(3)$_C$ is 
 \bea
 \mathcal{L} \supset N_g (c_u + c_d) T(q) \frac{g^2_s}{16\pi^2} \frac{a}{f_a} G \tilde{G},
 \eea 
with $T(q) = 1/2$ (the index of the fundamental representation of SU(3)$_C$) and the sum ($c_u + c_d$) being zero implies that the strong CP problem cannot be resolved, because the color anomaly coefficient ($N$) for $G\tilde{G}$ term is 0. Therefore, the lepton-specific DFSZ-like model can only work as an ALP model.

The convention of $c_{u} /f_a, c_{d} /f_a$ and $c_{\ell}/ f_a$ can be expressed as $g^{\text{eff}}_{a\bar{q}{q}}$, $-g^{\text{eff}}_{a\bar{q}{q}}$ 
and $g^{\text{eff}}_{a\bar{\ell}\ell}$ respectively. Thus, eq.~(\ref{eq:dfszlikegammamugamma5FFtilde}) could be rewritten as follows: 
\bea \label{eq:Lag_gmug5}
\mathcal{L} &\supset g^{\text{eff}}_{a\bar{q}q}  \displaystyle{}\frac{\partial_\mu a}{2} \bar{u}\gamma^\mu\gamma_5 u -  g^{\text{eff}}_{a\bar{q}q}  \frac{\partial_\mu a}{2} \bar{d}\gamma^\mu\gamma_5 d +	g^{\text{eff}}_{a\bar{\ell}\ell} \frac{\partial_\mu a}{2} \bar{\ell}\gamma^\mu\gamma_5 \ell \non\\
		&\displaystyle{}-(N^q_c Q^2_u g^{\text{eff}}_{a\bar{q}q} -N^q_c Q^2_d g^{\text{eff}}_{a\bar{q}q} + N^{\ell}_c Q^2_{\ell} g^{\text{eff}}_{a\bar{\ell}\ell} )\frac{e^2}{16\pi^2}a F_{\mu\nu}\tilde{F}^{\mu\nu},
\eea

In the lepton-specific DFSZ-like model, the effective ALP-photon coupling comes from the chiral anomaly and also can arise from the ALP-fermion coupling at 1-loop, as shown in the left panel of figure~\ref{fig:axionloopeepair}. The 1-loop contribution for $g_{a\gamma\gamma}^{\text{1-loop}}$ is \cite{Bauer:2017ris,Spira:1995rr}
\bea
\mathcal{L}^{\text{loop}}_{a\gamma\gamma} =  \bigg(N^q_c Q^2_u g^{\text{eff}}_{a\bar{q}q}  B(\tau_u) -N^q_c Q^2_d g^{\text{eff}}_{a\bar{q}q}  B(\tau_d) + N^{\ell}_c Q^2_{\ell} g^{\text{eff}}_{a\bar{\ell}\ell}  B(\tau_\ell) \bigg)\frac{e^2}{16\pi^2}  a F_{\mu\nu}  \tilde{F}^{\mu\nu},
\eea
where $f = u,d,\ell$ and $B(\tau_f)$ are the loop functions given in eq.~(\ref{eq:notreelevlgaAcoupling}).
Therefore, adding the tree-level and 1-loop contribution together, the effective ALP-photon coupling can be expressed as follows:
\bea
g_{a\gamma\gamma}^{\text{eff}}&=&\frac{e^2}{16\pi^2}
		\left( \sum_{i=u,c,t} (B_1(\tau_i)-1)\frac{4}{3} g_{a\bar{q}q}^{\text{eff}} - \sum_{i=d,s,b}(B_1(\tau_i)-1)\frac{1}{3} g_{a\bar{q}q}^{\text{eff}} \right) \non\\ &+& \frac{e^2}{16\pi^2} \left(\sum_{i=e,\mu,\tau} (B_1(\tau_i)-1)g_{a\bar{\ell}\ell}^{\text{eff}} \right),
\eea
 where $B(\tau_f)$ is expressed as follows:
\bea \label{eq:notreelevlgaAcoupling}
B(\tau_f)  = 1 -\tau_f f^2(\tau_f),  \,\, \tau_f &\equiv&  \frac{4 m^2_f}{m^2_a},\,\, f(\tau_f) =
\left \{
\begin{array}{cc}
\displaystyle{} \text{arcsin} \frac{1}{\sqrt{\tau_f}} ; & \tau_f \geq 1 \\
 \displaystyle{}\frac{\pi}{2} + \frac{i}{2} \text{ln}\frac{1+\sqrt{1-\tau_f}}{1-\sqrt{1-\tau_f}}; &  \tau_f < 1
\end{array} 
\right.,
\eea
where $f$ are the SM fermions. The loop functions have two useful limits that, 
\bea
\tau_f  \to   \infty , B(\tau_f) \approx -\frac{m^2_a}{12 m^2_f} ; \quad
\tau_f \to 0,  B(\tau_f)  \approx 1,
\eea
Therefore, for $m_f \gg m_a$ the 1-loop contribution for ALP-photon couplings are suppressed, but are comparable to chiral rotation contributions for $m_f \lesssim m_a$. 

  \begin{figure}[t!]
   	\centering
   	\includegraphics[width=0.6\linewidth]{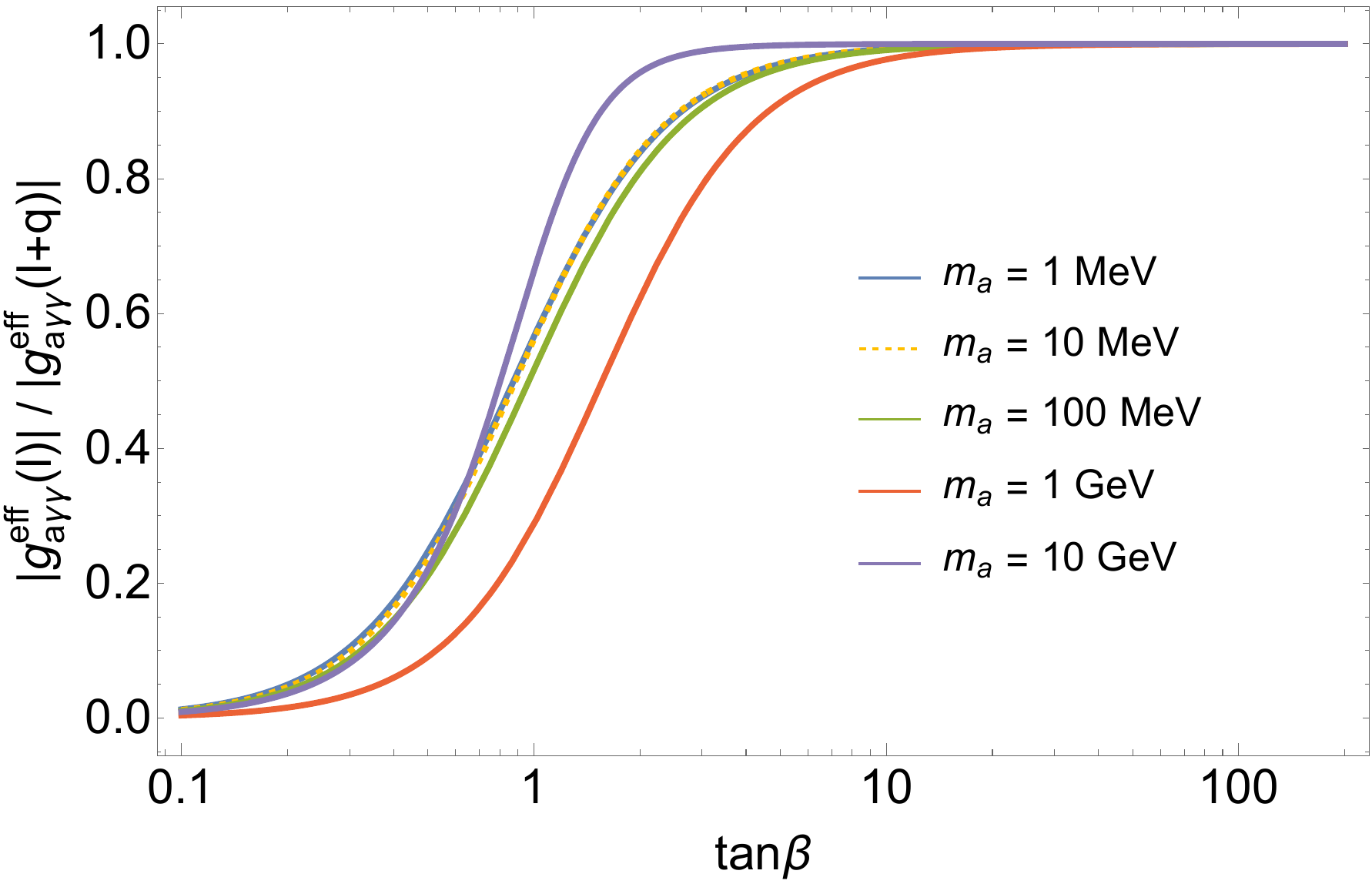} 
   	\caption{The ratio between the contribution to $g_{a\gamma\gamma}^{\text{eff}}$ from only leptons and contribution to $g_{a\gamma\gamma}^{\text{eff}}$ from quarks and leptons. We choose benchmark ALP mass to be $m_a=1$ MeV (blue line), $m_a=10$ MeV (yellow dashed line), $m_a=100$ MeV (green line), $m_a=1$ GeV (red line) and $m_a=10$ GeV (purple line). 
    }
   	\label{fig:dfsz}
   \end{figure}
   
In this study, we are going to focus on the ALP-lepton couplings. Contribution to ALP-photon couplings with leptons only are in the same level with both quarks and leptons in the regime of large $\tan\beta$ where $v_1 \gg v_2$. Numerically, we found that for $\tan\beta > 10$, the contributions from all fermions and leptons only are of similar magnitude where we show their ratio in figure~\ref{fig:dfsz}. In the general DFSZ model, the values of $\tan\beta$ are constrained by the perturbative unitarity bound, which lies between 0.25 and 170~\cite{DiLuzio:2020wdo, Bjorkeroth:2019jtx, DiLuzio:2016sur, DiLuzio:2017chi}. Therefore, we
choose $\tan\beta>10$ in this study and neglect the ALP-quark couplings and its contribution to ALP-photon coupling. 
This is a good approximation for examining the concurrence features of ALP-photon coupling and ALP-electron coupling. Other physical scalars are quite heavy to avoid phenomenological constraints, thus the lower-energy phenomenon are the physical ALP in the scalar sector. We take the ratio between the electron-ALP coupling ($g^{\text{eff}}_{a\bar{e}e}$) and photon-ALP coupling with leptons only ($g^{\text{eff}}_{a\gamma\gamma}$), and the expression is 
\bea \label{eq:caAAovercaee}
 \displaystyle{}\frac{g^{\text{eff}}_{a\gamma\gamma}}{g^{\text{eff}}_{a\bar{e} e}} &=& \frac{\sum_{\ell= e,\mu,\tau}  \left( B(\tau_\ell) - 1\right)  \frac{N_c Q^2_f e^2}{16\pi^2} g^{\text{eff}}_{a\bar{\ell}\ell}}{g^{\text{eff}}_{a\bar{e}e}}  \nonumber \\
&=& 
\frac{\alpha_{\text{QED}}}{4\pi}
\frac{ \left(g^{\text{eff}}_{a\bar{e}e} \left( B(\tau_e) - 1\right) + g^{\text{eff}}_{a\bar{\mu}\mu} \left( B(\tau_\mu) - 1\right) + g^{\text{eff}}_{a\bar{\tau}\tau} \left( B(\tau_\tau) - 1\right) \right)
}{ g^{\text{eff}}_{a\bar{e}e}} .
\eea

Since we have the same $g^{\text{eff}}_{a\bar{\ell}\ell} $ for all three leptons, the ratio above can be simplified as $|\frac{\alpha_{\text{QED}}}{4\pi} \sum_{\ell= e,\mu,\tau} \left( B(\tau_f) - 1 \right)|$.  If the ALP only couples to electrons, the ratio factor can be simplified to $\frac{\alpha_{\text{QED}}}{4\pi} \left(B(\tau_e) - 1\right)$. Therefore, the ratio factor serves as an effective model constraint when considering the concurrent effects of two couplings. For our analysis, we adopt $m_e = 0.000511$ GeV, $m_\mu = 0.1056$ GeV, and $m_\tau = 1.78$ GeV. 

\begin{figure}[t!]
    \centering
     \begin{subfigure}
    {
        \hspace{-0.8cm}
        \includegraphics[scale=.3]{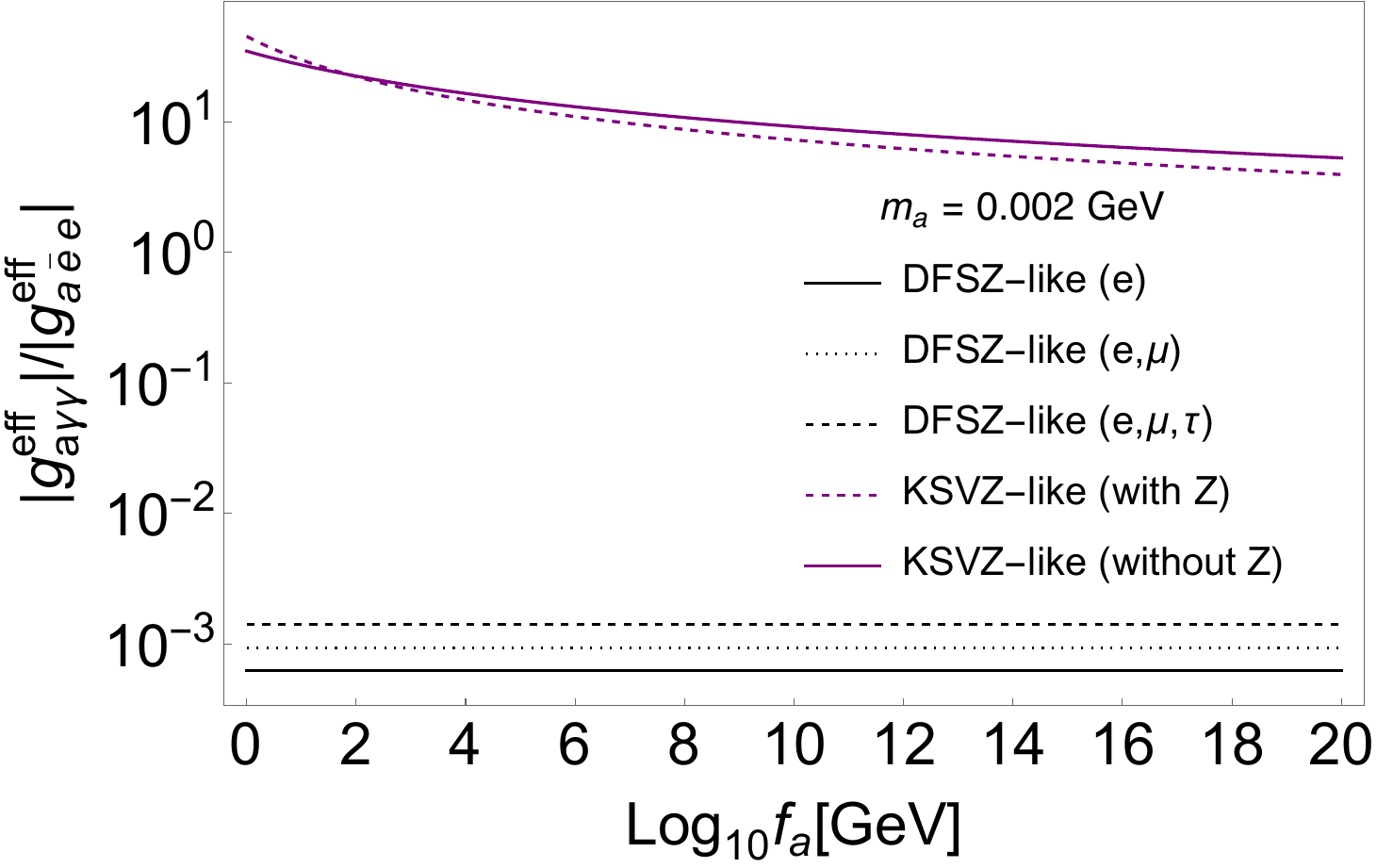}\hspace{1cm}
         \includegraphics[scale=.3]{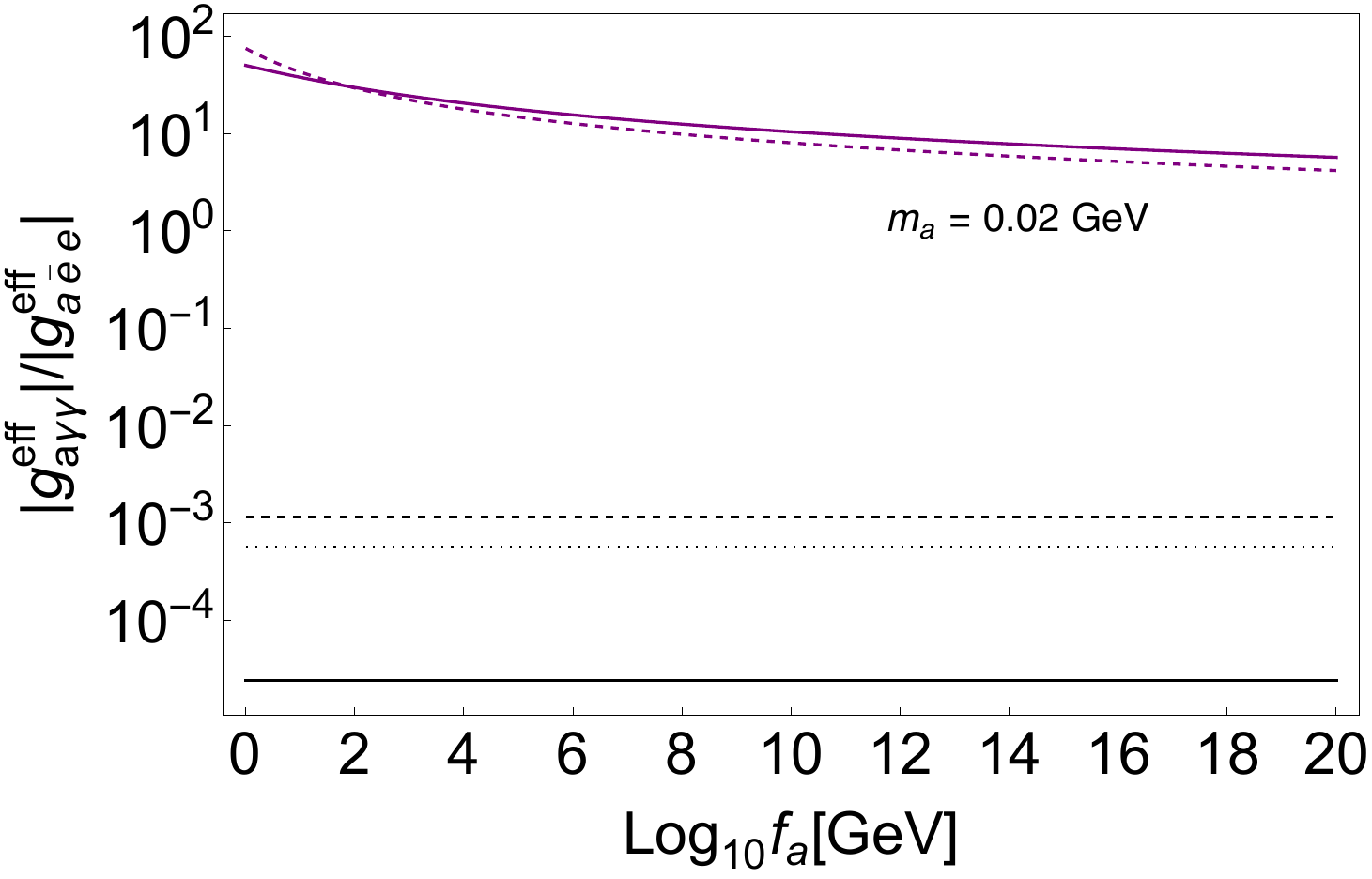}
    }
    \subfigure
   {
         \hspace{-0.8cm}
         \includegraphics[scale=.3]{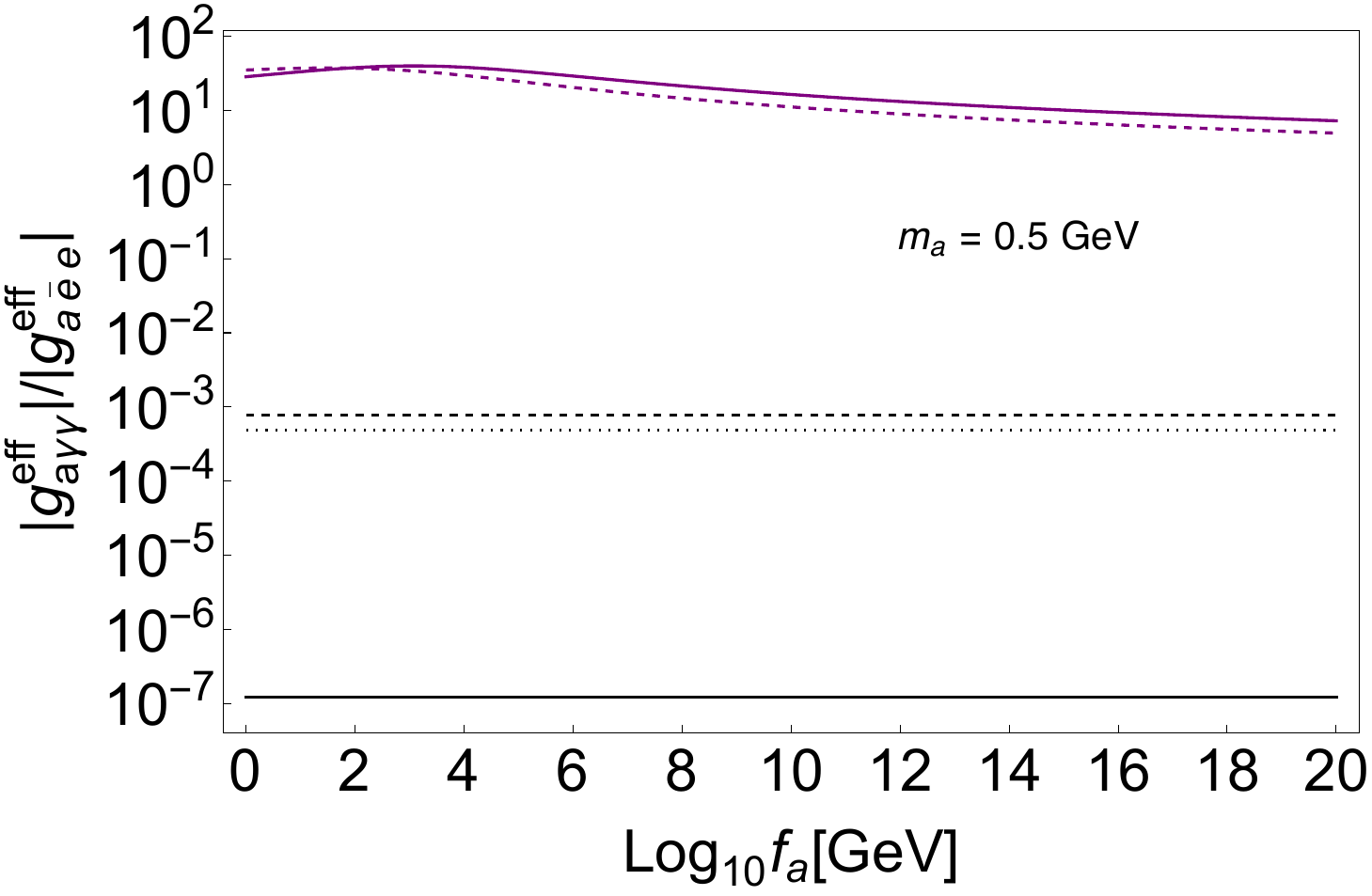}\hspace{1cm}
         \includegraphics[scale=.3]{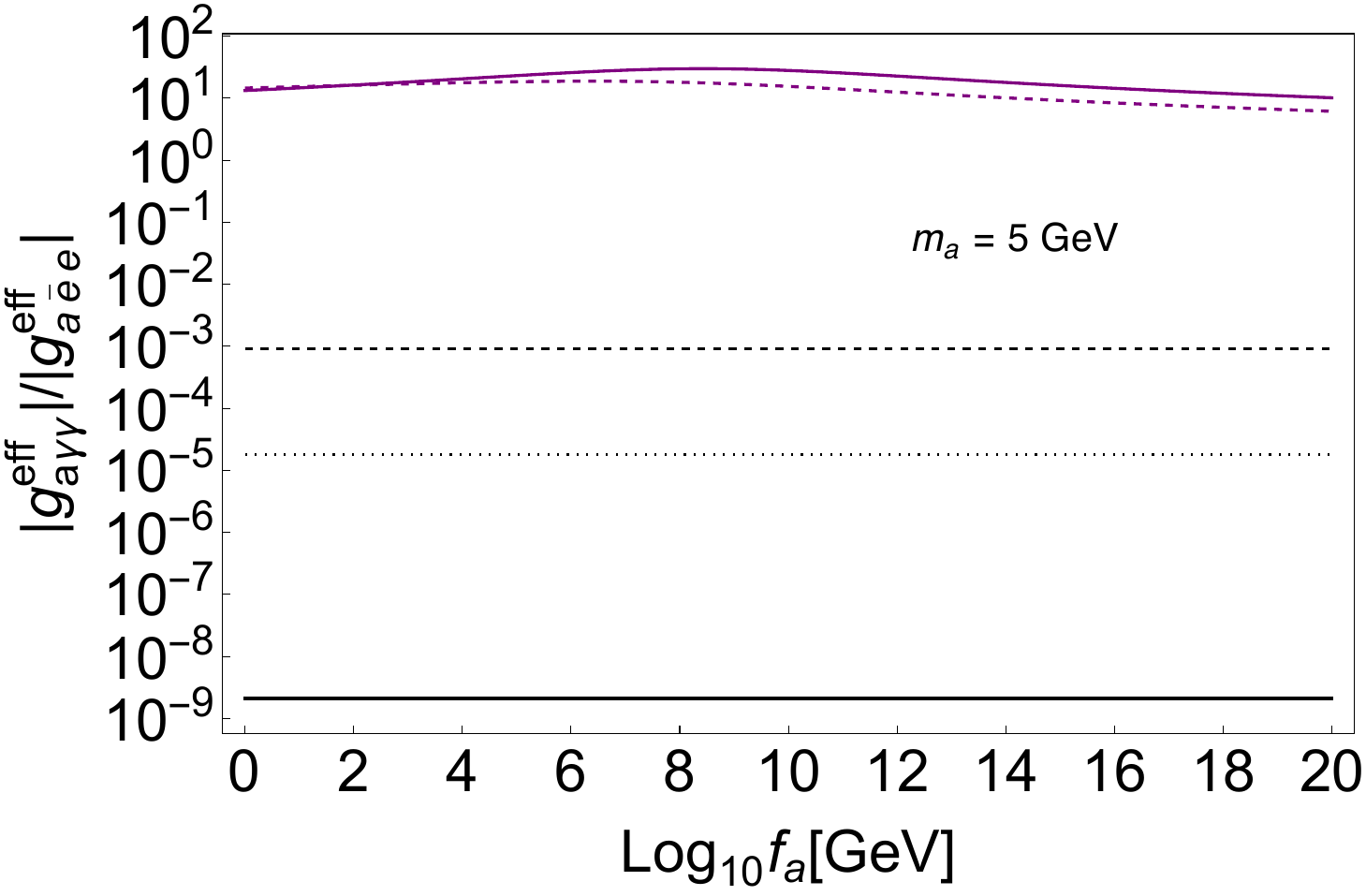}
    }
    \caption{ $|g^{\text{eff}}_{a\gamma\gamma}|/|g^{\text{eff}}_{a\bar{e} e}|$ as a function of $f_a$ (in GeV) from two UV models with $m_a$: 0.002 GeV (top left), 0.02 GeV (top right), 0.5 GeV (bottom left), and 5 GeV (bottom right). In the figure, the KSVZ-like model is represented by purple lines, where the solid line indicates contributions to $g^{\text{eff}}_{a\bar{e} e}$ from $F\tilde{F}$ only, and the dashed line includes contributions from $F\tilde{Z}$ and $Z\tilde{Z}$. The DFSZ-like model is represented by black lines, with $g^{\text{eff}}_{a\gamma\gamma}$ contributions from only $e$ loops indicated by a solid line, from $e,\mu$ loops by a dotted line, and from $e,\mu,\tau$ loops by a dashed line.
    }\label{fig:UVratiowithMa}
    \end{subfigure}	
\end{figure}

\begin{figure}[t!]
    \centering
     \begin{subfigure}
    {
         \includegraphics[scale=.3]{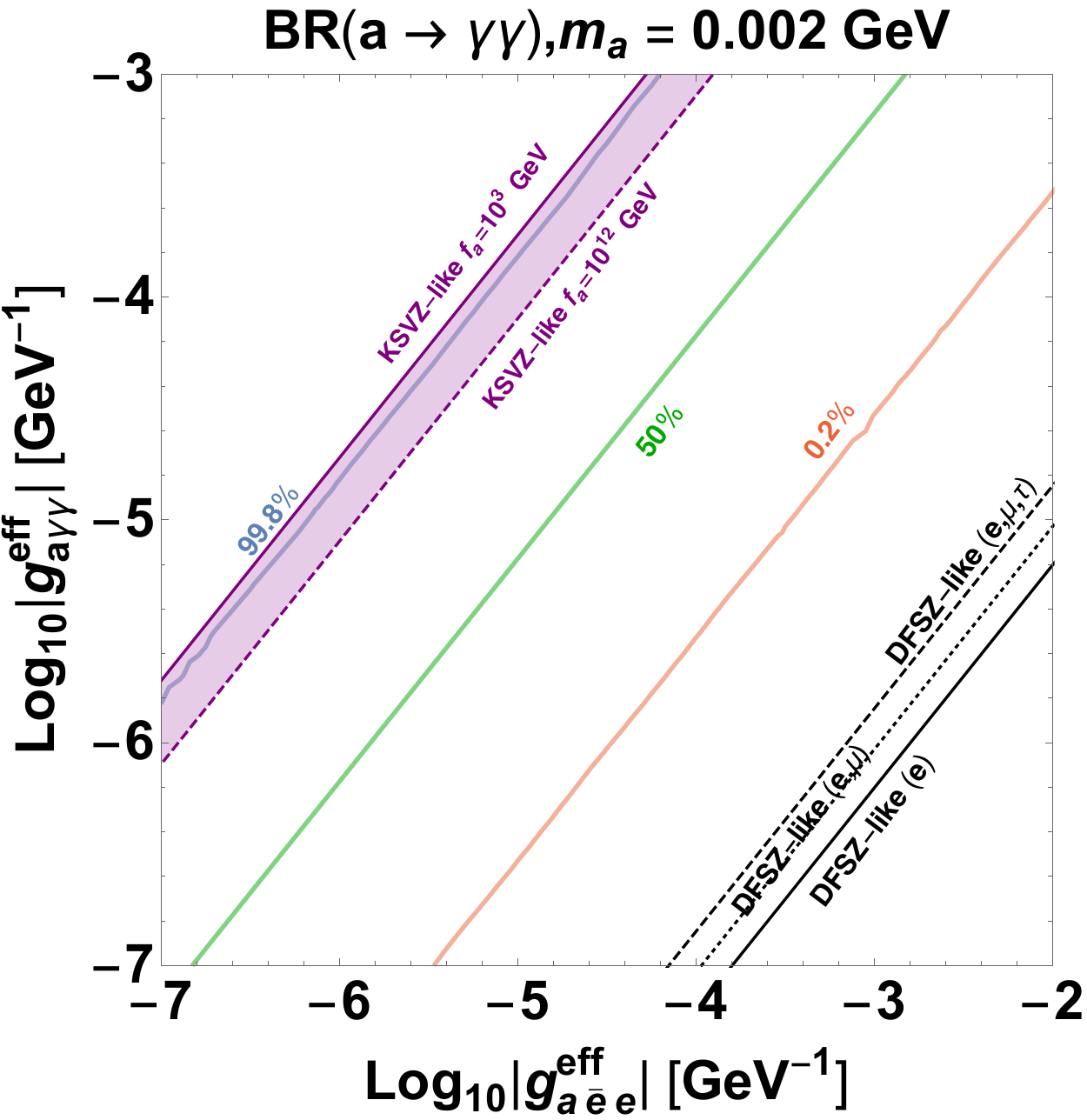}\hspace{1cm}
         \includegraphics[scale=.3]{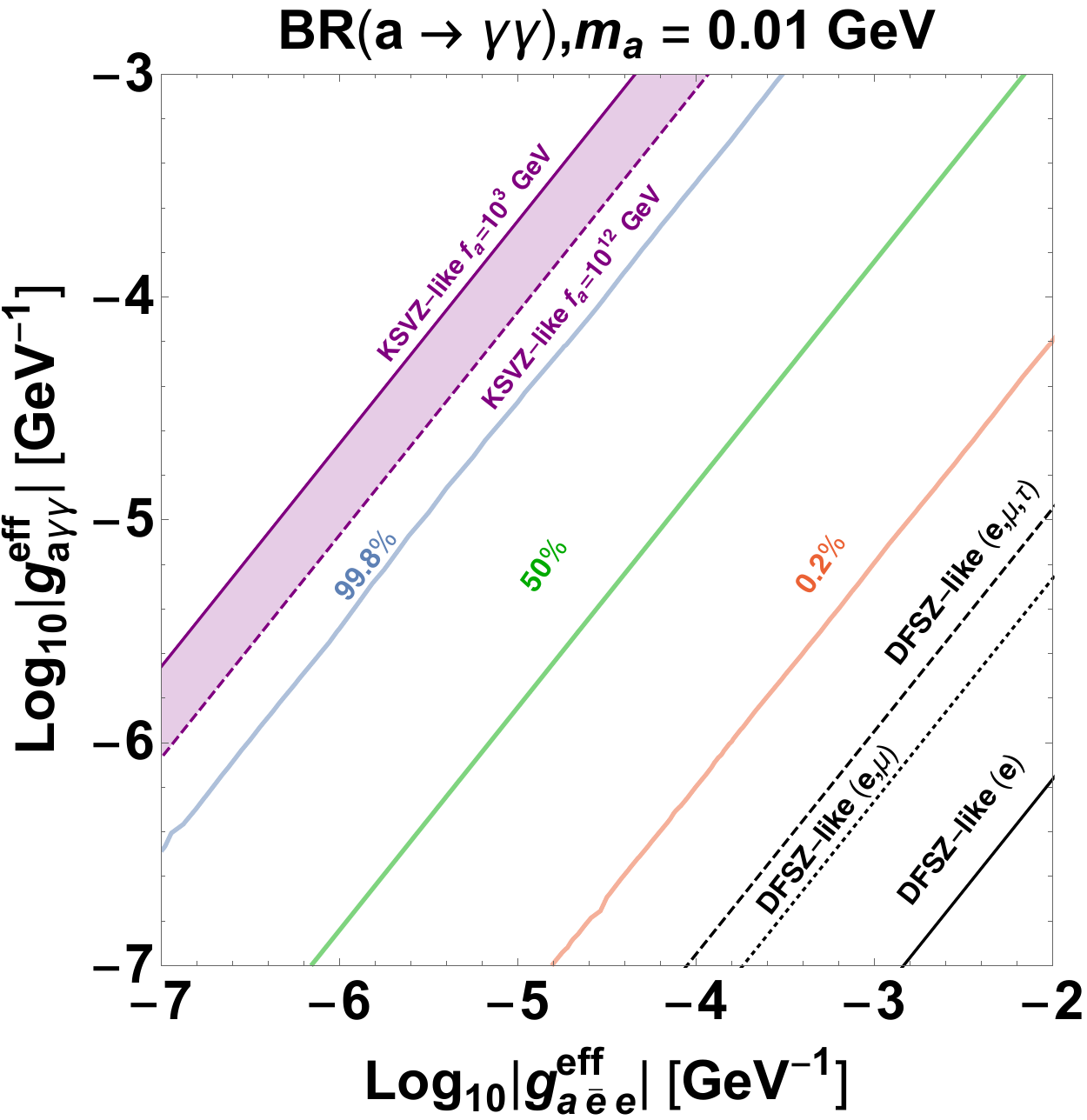}
    }
    \subfigure
   {
         \includegraphics[scale=.3]{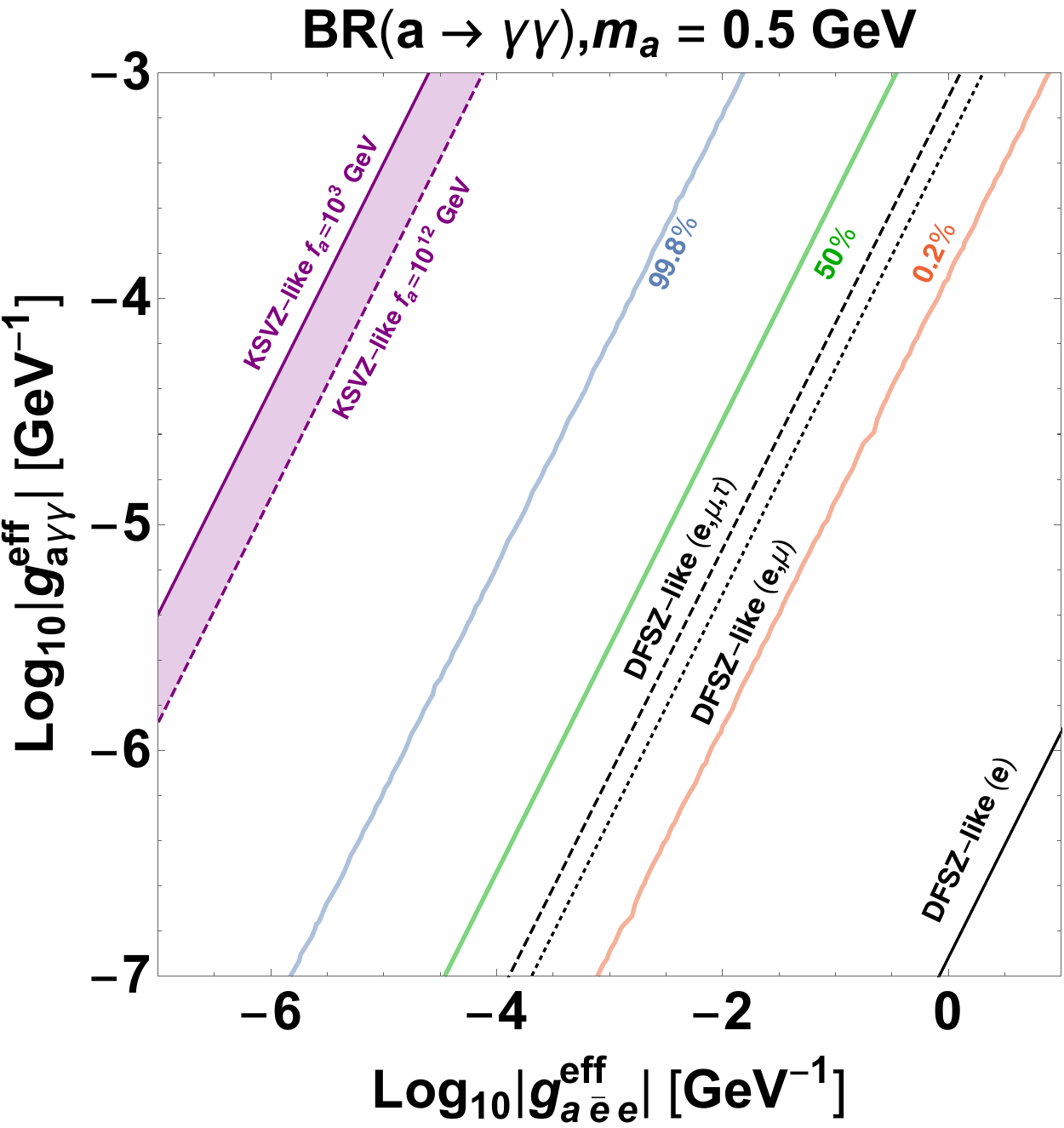}\hspace{1cm}
         \includegraphics[scale=.3]{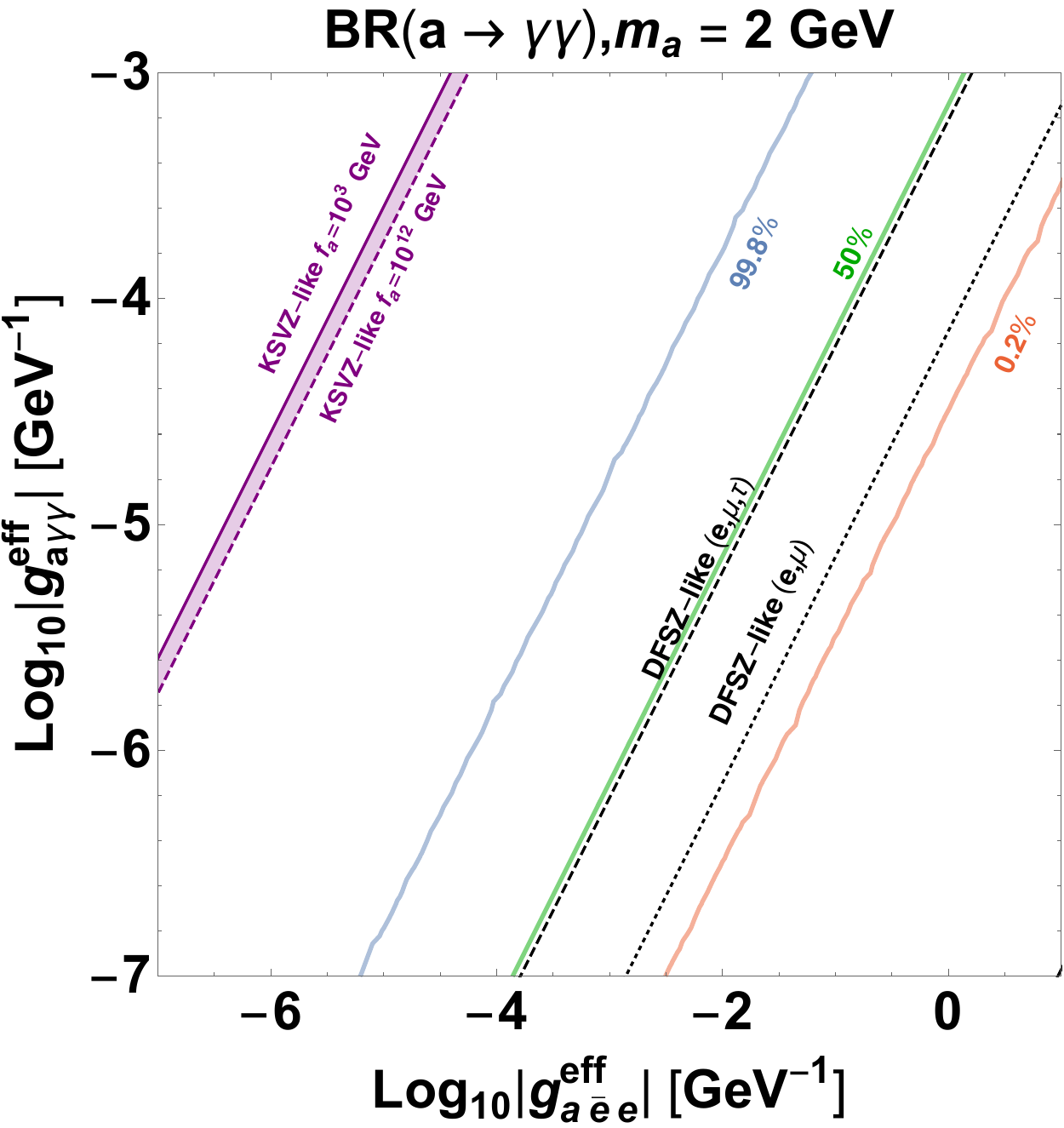}
    }
    \caption{ BR($a\to \gamma\gamma$) (solid blue for 99.8\%, solid green for 50\% and solid orange for 0.2\%) in the  [$|g^{\text{eff}}_{a\bar{e} e}|,|g^{\text{eff}}_{a\gamma\gamma}|$] plane with $m_a$: 0.002 GeV (top left panel),0.01 GeV (top right panel), 0.5 GeV (bottom left panel) and 2 GeV (bottom right panel). Purple bands are allowed KSVZ-like model with $f_a = 10^3$ to $10^{12}$ GeV. The black lines are the DFSZ-like photon coupling produced from $e$ loop only(solid), $ e,\mu$ loops(dotted), and $e,\mu, \tau$ loops(dashed) available region.
    }\label{fig:UVBRwithMa}
    \end{subfigure}	
\end{figure}
\subsection{Model interpretations}

This section discusses the decay properties of ALP with respect to electrons $|g^{\text{eff}}_{a\bar{e}e}|$ and photons $|g^{\text{eff}}_{a\gamma\gamma}|$, and their characteristics in the context of the aforementioned ultraviolet models. The combined effects of these two couplings and the mass of ALP $m_a$ result in an increased decay width of the ALP and are crucial for determining the branching ratios.
In figure~\ref{fig:UVratiowithMa}, the ratio $|g^{\text{eff}}_{a\gamma\gamma}|/|g^{\text{eff}}_{a\bar{e}e}|$ from two ultraviolet models are plotted as a function of $f_a$ for four values of $m_a$: 0.002, 0.02, 0.5, and 5 GeV, from top left to bottom right. The KSVZ-like model ratio (purple line) follows eq.~(\ref{eq:ksvzcaAAovercaee}). The DFSZ-like model ratio follows eq.~(\ref{eq:caAAovercaee}), represented by three black lines in the figure (the solid line corresponds to only $e$ in the fermion loop, the dotted line includes $e$ and $\mu$, and the dashed line consists of all three leptons, $e$, $\mu$, and $\tau$). The ratio of the DFSZ-like model is independent of $f_a$ because the coupling decay constants $f_a$ are cancelled out, while the KSVZ model is dependent and correlated with two parameters, $m_a$ and $f_a$. Due to the modulus of real and imaginary contribution, the maximum ratio would vary with values of $f_a$ and ranges from $10^{0.5}$ to $10^{2}$ for $f_a$ from $10^0$ to $10^{20}$ GeV, which changes very mild due to logarithmic dependence. The values of the ratio between the two couplings in the KSVZ-like model are consistently greater than those in the DFSZ-like model for all four choices of $m_a$. This disparity can be attributed to the presence of the correlation factor $\alpha_{\text{QED}}$ within the effective ALP-electron coupling, as described by eq.~(\ref{eq:KSVZinducedaeefromaAA}). This factor enhances the ratio by reducing the denominator to a significantly smaller value, resulting in a stronger preference for ALP-photon coupling in the KSVZ-like model.

In figure~\ref{fig:UVBRwithMa}, we show four plots with BR($a\to \gamma\gamma$) (solid blue for 99.8\%, solid green for 50\% and solid orange for 0.2\%) in the $\{|g^{\text{eff}}_{a\bar{e} e}|,|g^{\text{eff}}_{a\gamma\gamma}| \}$ 2D plane with $m_a$ equals to 0.002 GeV (top left panel), 0.01 GeV (top right panel), 0.5 GeV (bottom left panel) and 2 GeV (bottom right panel) respectively. The decay branching ratio to electrons, BR($a \to \bar{e} e$), on the other hand, will then equal $1$ - BR($a\to \gamma\gamma$). We illustrated the ultraviolet models, represented by the purple band for KSVZ-like and black lines (solid for $e$ contribution only, dotted for $e, \mu$ contributions, and dashed for all three lepton, $e,\mu,\tau$ contributions) for DFSZ-like. We found that BR($a \to \gamma\gamma$) is dominant when both $g^{\text{eff}}_{a\gamma\gamma}$ and $g^{\text{eff}}_{a\bar{e} e}$ couplings are in the same order for heavier $m_a$ states (see plots with $m_a =$ 0.5 and 2 GeV). Conversely, when $m_a$ decreases, BR($a\to \bar{e} e$) increases because that the decay ratio between $a\to \gamma \gamma$ and $a \to \bar{e} e$ is directly proportional  to $m^2_a/8 m^2_e$. Therefore, these decay properties can be combined with existing experimental results to provide testable predictions.

\section{Status of ALP searches at colliders and beamdump experiments}\label{sec: experiment results}

Based on various ALP masses, ALP searches have been conducted in different experiments. In ultralight and light mass regions ($m_a \ll  m_e$), ALP can be probed via astronomy and cosmology searches. In the heavy mass regions, high energy colliders have reached several hundred GeVs (several ALP productions are even correlated with a gauge or Higgs bosons)~\cite{Brivio:2017ije, Bauer:2017ris, Bauer:2018uxu}. This work focuses on an intermediate mass range, between $2 m_e$ and up to 10 GeV. The main interest lies in the electron-ALP and photon-ALP couplings, where ALPs are produced through electron-positron annihilations, electron collisions in the fixed target or electron beam dump experiments.

\subsection{Electron positron collider experiments}
For the constraint on the ALP-photon coupling, the limit from the Belle-II experiment \cite{Belle-II:2020jti} is considered. This limit covers the ALP mass detection range between 0.2 and 9.7 GeV from electron-positron collisions ($e^++e^- \to \gamma + a$) followed by $a\to \gamma\gamma$, with three photons in the final state. The data was collected with an integrated luminosity of 445 pb$^{-1}$ at the center-of-mass energy of $\sqrt{s} = 10.58$ GeV. The background is dominated by  $e^+e^- \to e^+e^-(\gamma)/,\gamma\gamma(\gamma)$ and pseudoscalar meson processes, such as $\pi^0, \eta,\eta'$ with $\gamma$ in the final state. For ALP-electron coupling constraint, the exclusion limit comes from BaBar \cite{Batell:2009yf, BaBar:2014zli, Bauer:2017ris, Bauer:2018uxu} in the mass region roughly between 0.02 and 9.2 GeV. The data collection has been performed within a luminosity of 514 fb$^{-1}$ at $\Upsilon (2S, 3S, 4S)$ resonance. The background events were simulated similarly to the Belle-II search, while BaBar used electron-positron collisions with two leptons and one photon as the final products.

\begin{figure}[t!]
\hspace{1cm}
        \includegraphics[scale=.75]{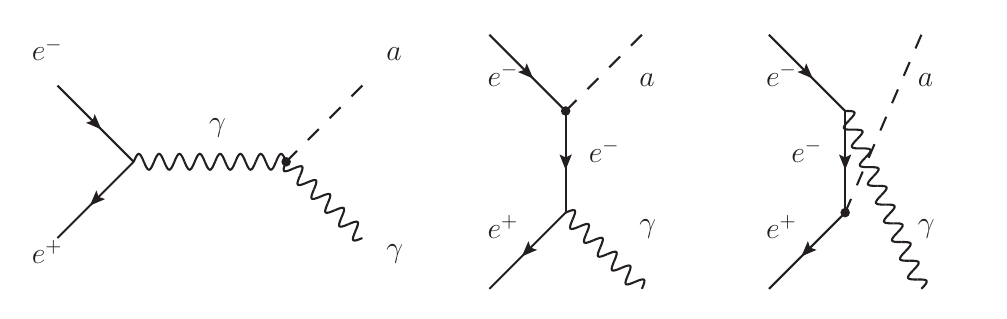}
    \caption{  Feynman diagrams for production of ALP ($a$) from electron-positron collider. The left channel is the s-channel, the middle one is the t-channel, and the right diagram is the u-channel.
    } \label{fig:eetoaA}
\end{figure}

The tree-level Feynman diagrams in figure~\ref{fig:eetoaA} are associate ALPs production with ALP-photon and ALP-electron couplings in the collider searches. The analytical formula for the cross-section of $e^+e^-\to a\gamma$ can be obtained by integrating the differential cross-section concerning the scattering angle, as calculated using the programs FeynRules 2.0~\cite{Alloul:2013bka}, and FeynCalc 9.3~\cite{Mertig:1990an, Shtabovenko:2016sxi, Shtabovenko:2020gxv} with patched FeynArts 3.11~\cite{Hahn:2000kx}, and it can be expanded using small $m_e$ as follows,
\bea \label{eq:eetoALPAme5}
\sigma_{e^+e^-\to a\gamma} &=&\alpha _{\text{qed}} \bigg[ \frac{ (g^{\text{eff}}_{a\gamma\gamma})^2  \left(s-m_a^2\right)^3}{24 s^3} \nonumber\\&+&
m^2_e
  \left( \frac{  (g^{\text{eff}}_{a\gamma\gamma})^2 \left(m_a^2-s\right)^4+(3 \ln{4})g^{\text{eff}}_{a\gamma\gamma} g^{\text{eff}}_{a\bar{e}e} s  \left(m_a^2-s\right)^3}{6 s^4 \left(s-m_a^2\right)}\right) \nonumber\\&+&
m^2_e
  \left( \frac{(3\ln{4}) (g^{\text{eff}}_{a\bar{e}e})^2 s^2  \left(m_a^4+s^2\right)}{6 s^4 \left(s-m_a^2\right)} \right)
\bigg] +\mathcal{O} (m^4_e) .
\eea 
The first term in the first line, $(g^{\text{eff}}_{a\gamma\gamma})^2 $ term in eq.~(\ref{eq:eetoALPAme5}), is the main term for the ALP-photon coupling in the $e^+e^-\to a \gamma$ cross-section process which corresponds to the $s-$channel diagram in the left panel of figure~\ref{fig:eetoaA}.  If $g^{\text{eff}}_{a\gamma\gamma}$ and $g^{\text{eff}}_{a\bar{e}e}$ are similar in magnitude, the ALP-photon coupling has a greater impact on the cross-section for low $m_a$ values, and it is because the interference term $g^{\text{eff}}_{a\gamma\gamma}g^{\text{eff}}_{a\bar{e}e}$ and ($g^{\text{eff}}_{a\bar{e}e}$)$^2$ is suppressed by $m^2_e/s$.

\begin{figure}[t!]
    {\hspace{0.5cm}
     \includegraphics[scale=.3]{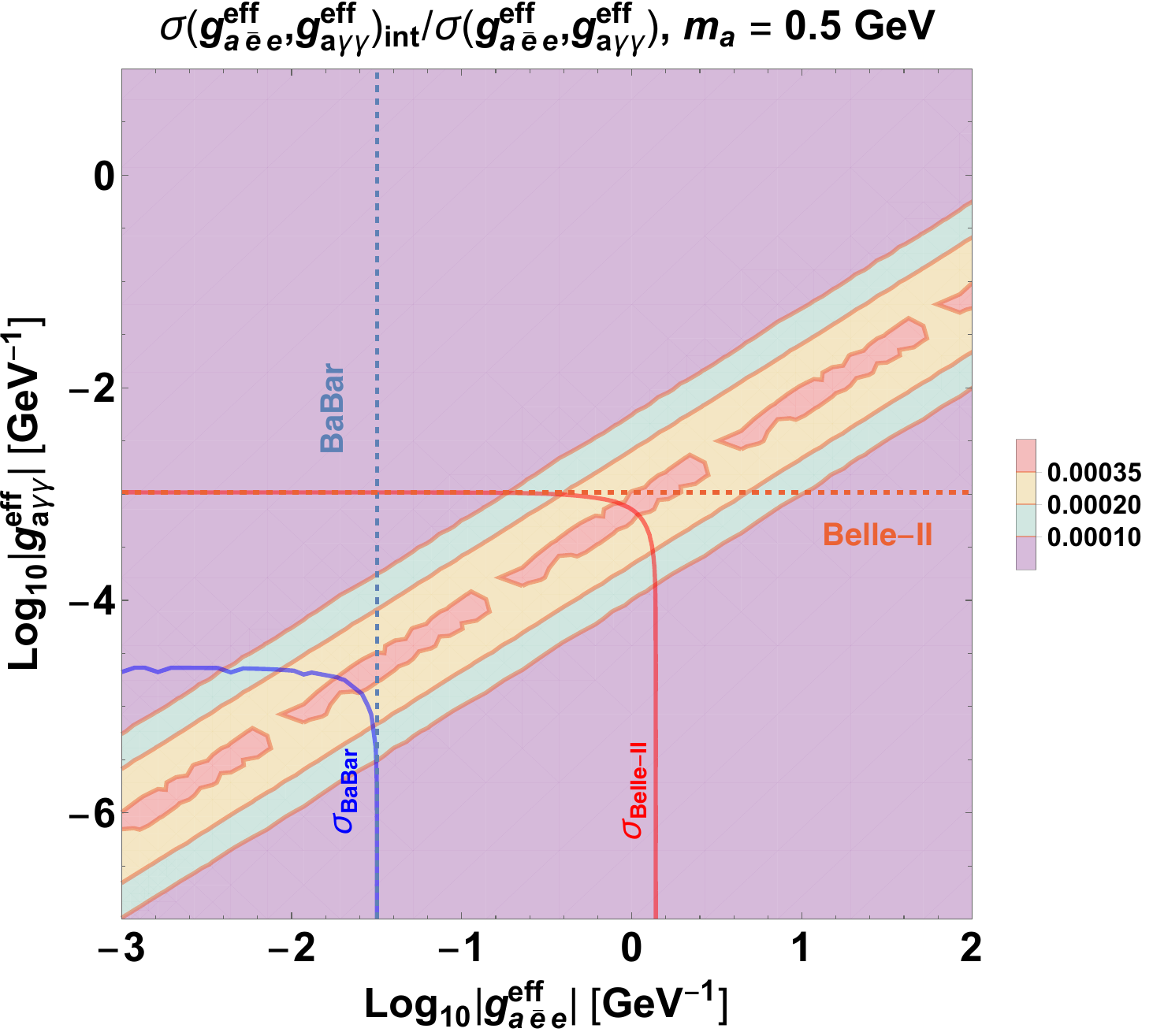}
        \includegraphics[scale=.3]{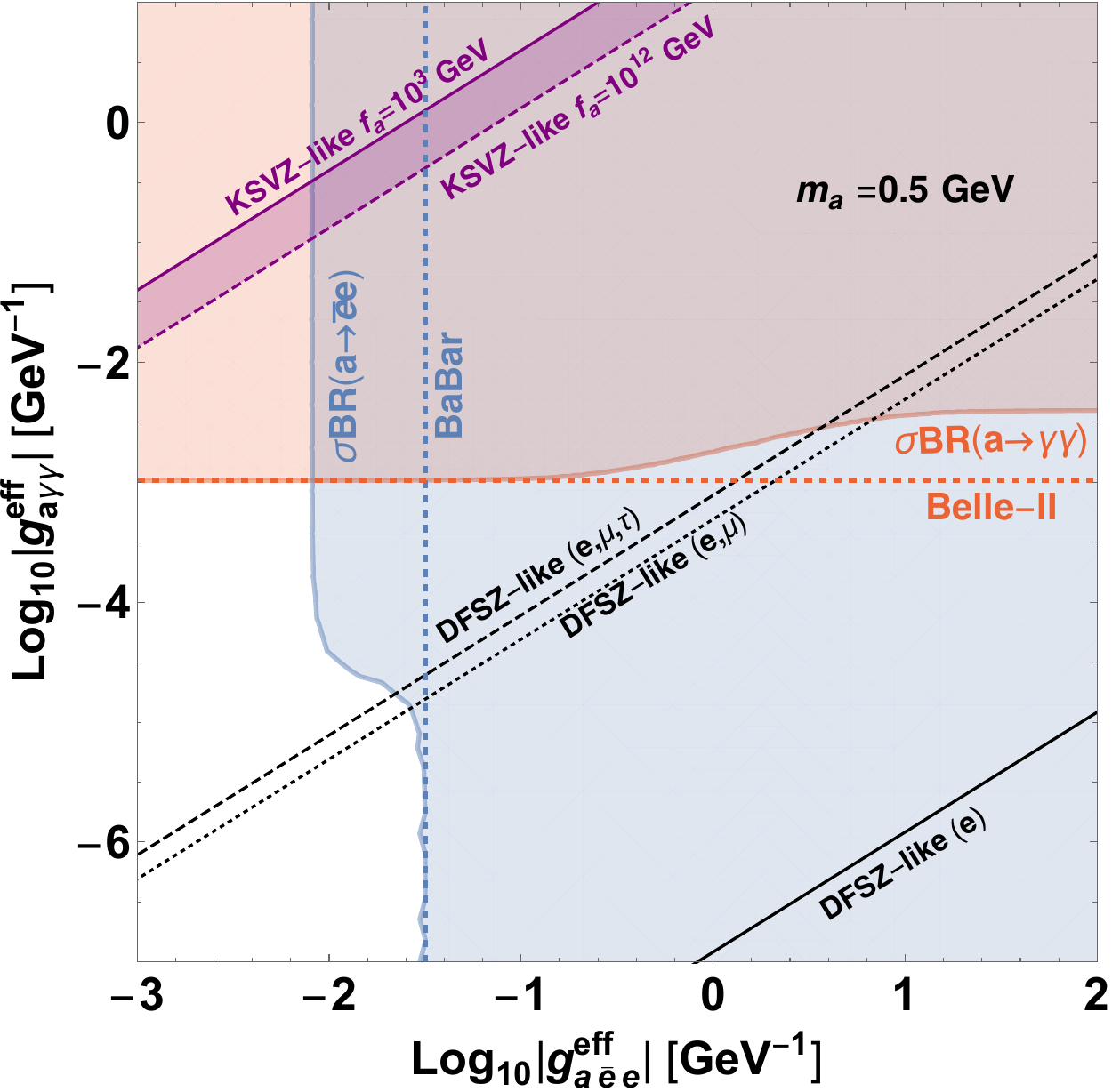}
    }
    \caption{Left panel: $\sigma_{\text{int}}/ \sigma_{\text{total}} $ in  [$|g^{\text{eff}}_{a\bar{e}e}|$,$|g^{\text{eff}}_{a\gamma\gamma}|$] plane with rescaled $\sigma$ only (Babar limit in blue line and Belle-II limit in red line).   Right panel: The exclusion region based on the product of $\sigma$ with BR($a\to \bar{e}e/\gamma\gamma$) (light blue/orange) for ALP mass $m_a = 0.5$ GeV. The region on the right of vertical light blue dashed line corresponds to the existing BaBar limit on ALP-lepton coupling, while the region above the horizontal orange dashed line represents the existing limit on ALP-electron coupling from Belle-II data. The black diagonal lines are the DFSZ-like photon coupling produced from $e$ loop only(solid), $ e,\mu$ loops(dotted), and $e,\mu, \tau$ loops(dashed) available region. Purple bands are allowed KSVZ-like  region with $f_a = 10^3$ to $ 10^{12}$ GeV. 
    } \label{fig:eegaegaAMa0d5}
    \end{figure}

The constraints set by the Belle-II~\cite{Belle-II:2020jti} and BaBar~\cite{BaBar:2014zli} searches covering single coupling, can be adapted to our concurrent scenario. In the left panel of figure~\ref{fig:eegaegaAMa0d5}, we depict the $\sigma (g^{\text{eff}}_{a\bar{e}e},g^{\text{eff}}_{a\gamma\gamma})_{\text{int}}/ \sigma (g^{\text{eff}}_{a\bar{e}e},g^{\text{eff}}_{a\gamma\gamma}) $ ratio, where the numerator is the interference contribution in the cross-section and the denominator is the full cross-section. One can see that the ratio is as small as $\mathcal{O}(10^{-4})$, that the interference effect is negligible due to the $m_e^2 g^{\text{eff}}_{a\bar{e}e}/(s g^{\text{eff}}_{a\gamma\gamma})$ factor. In addition, we plot the contours of cross-section $\sigma$ equal to BaBar (solid blue) and Belle-II (solid red) limits to illustrate the excluded region in the 2D plane $[|g^{\text{eff}}_{a\bar{e}e}|, |g^{\text{eff}}_{a\gamma\gamma}|]$ at $m_a = 0.5$ GeV. 
For comparison, we have plotted the exclusion limits from previous experiments for single coupling scenario, indicated by the light blue and orange dashed lines for BaBar and Belle-II respectively. The regions on the upper-right side of solid curves would be ruled out. Compared to single coupling scenario, more regions are ruled out due to larger cross-section induced by concurrent couplings, for example, some areas on the left-hand side of the light blue dashed line.

To examine the cross-section of electron-positron annihilation with the ALP followed by di-electron and diphoton decay channels, we analyze the ALP final states for the existing limits by multiplying $\sigma (e^+ + e^- \to a + \gamma)$ with BR($a \to \bar{e}e / \gamma\gamma$). In the right panel of figure~\ref{fig:eegaegaAMa0d5}, with same benchmark point of $m_a = 0.5$ GeV, the product $\sigma \times $ BR($a \to \bar{e}e/ \gamma\gamma$) (light blue / orange exclusion regions) are plotted and only white area is allowed by experiments. The concurrent effect is induced by two factors. One is cross-section which reduces the allowed parameter space, and another is the branching ratio (BR), which 
in contrast, tends to relax constraints. For BaBar limits, We know from the left panel that bending from vertical to left is attributed to larger cross-section after considering $g_{a\gamma\gamma}^{\text{eff}}$, while bending from left to up is due to BR effect. Specifically, the region with $|g^{\text{eff}}_{a\bar{e}e}| <$ 10$^{-2}$ GeV$^{-1}$ and $|g^{\text{eff}}_{a\gamma\gamma}| >$ 10$^{-4.5}$ GeV$^{-1}$ evades the BaBar limit ($\sigma_{\text{BaBar}}$ on the left panel excludes $|g^{\text{eff}}_{a\bar{e}e}| >$ 10$^{-1.5}$ GeV$^{-1}$ and $|g^{\text{eff}}_{a\gamma\gamma}| >$ 10$^{-4.5}$ GeV$^{-1}$) because a large effective coupling $|g^{\text{eff}}_{a\gamma\gamma}|$ decreases the branching ratio BR($a\to \bar{e}e$), thereby suppressing the product of the cross-section and branching ratio ($\sigma$BR). For Belle-II limits, the open areas with $|g^{\text{eff}}_{a\bar{e}e}| >$ 10$^{-1}$ GeV$^{-1}$ and 10$^{-3} <|g^{\text{eff}}_{a\gamma\gamma}| < 10^{-2.5}$  GeV$^{-1}$ relax the bound by the suppression of BR($a\to \bar{e}e$).

\begin{figure}[t!]
    \centering
     \begin{subfigure}
    {
         \includegraphics[scale=.3]{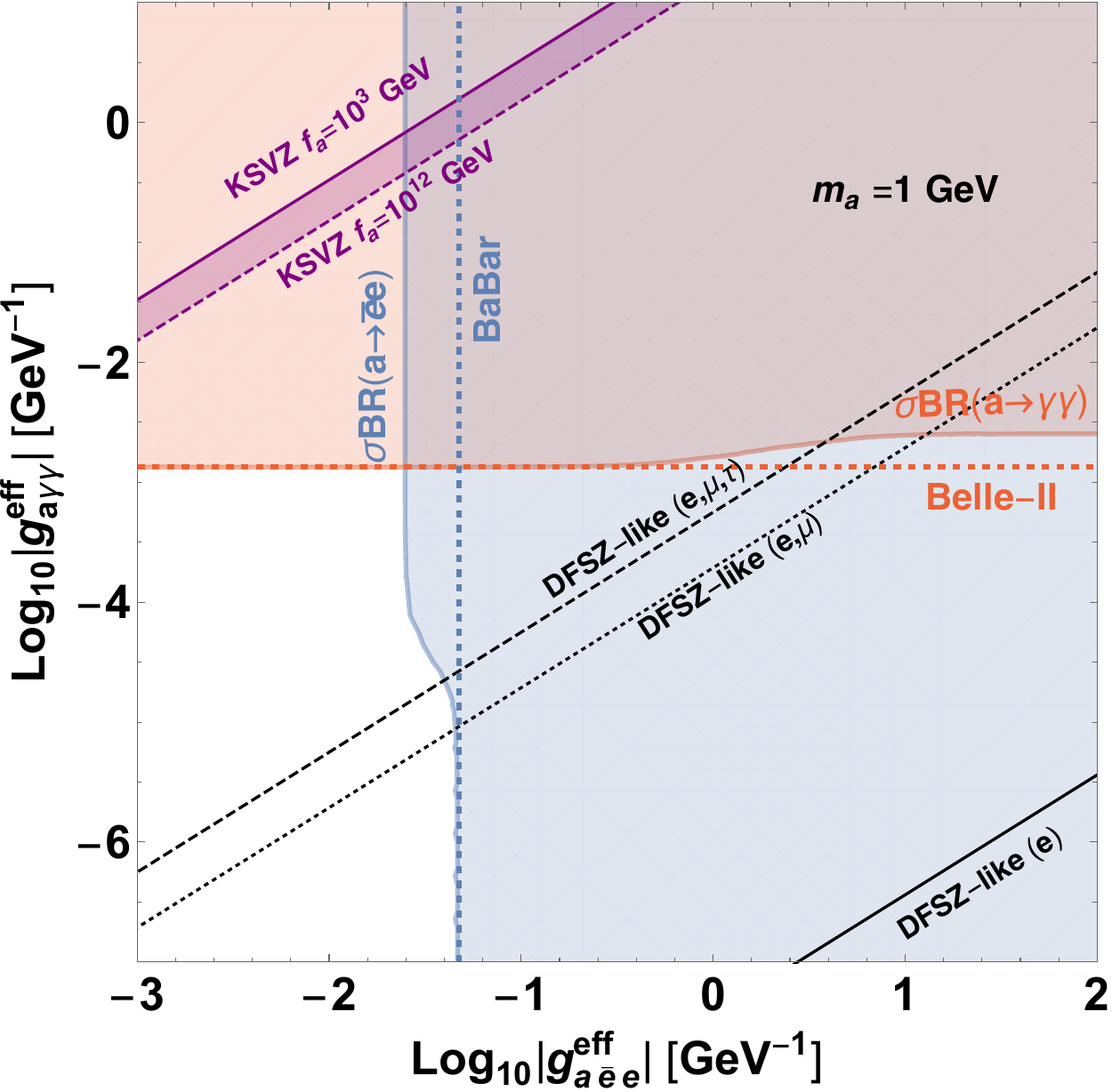}\hspace{1cm}
         \includegraphics[scale=.3]{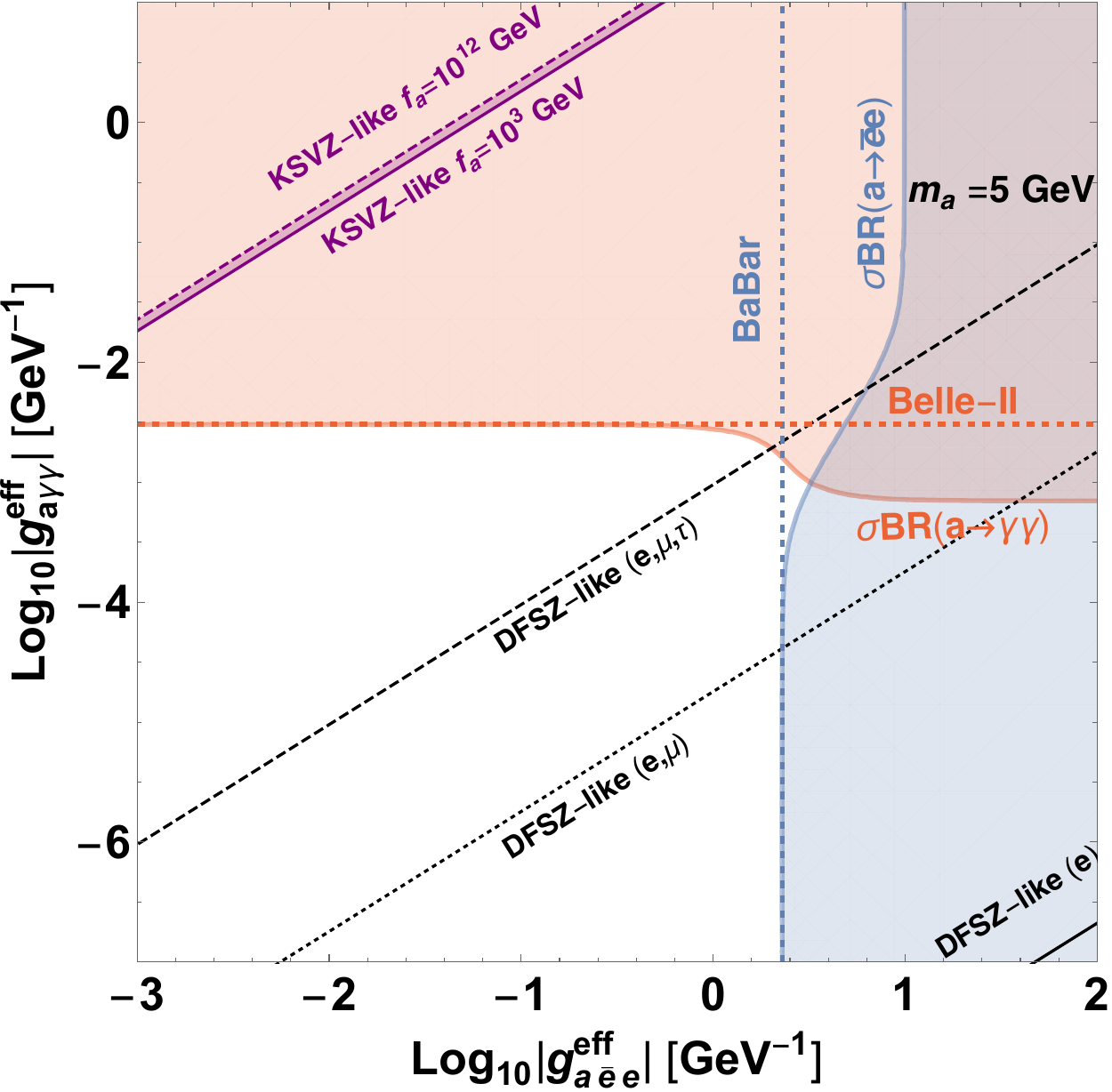}
    }
     \caption{ Same figure as right panel of figure~\ref{fig:eegaegaAMa0d5} with $m_a =$ 1 GeV (left panel) and 5 GeV (right panel). The white region is allowed. }\label{fig:gaegaAMa15}
    \end{subfigure}	
\end{figure}    

In the right panel of figure~\ref{fig:eegaegaAMa0d5}, we have incorporated UV model suggested parameter space from the modified DFSZ-like model with diagonal black lines (solid, dotted and dashed) and the KSVZ-like model with two purple lines (solid and dashed) corresponding to $f_a = 10^3$ GeV and $f_a = 10^{12}$ GeV, respectively. For $m_a = 0.3$ GeV, the KSVZ-like model is fully excluded 
in the 2D plane, while the DFSZ-like model are still survived for 
coupling to $(e,\mu)$ and $(e,\mu,\tau)$ but the electron only coupling is excluded.

We consider more ALP masses, $m_a = 1$ GeV and $m_a = 5$ GeV, in the left and right panels of figure~\ref{fig:gaegaAMa15}. The figure for $m_a = 1$ GeV is qualitatively similar to $m_a = 0.5$ GeV, but the allowed region expands. For $m_a = 5$ GeV, comparing with the single coupling scenario, the BaBar constraint becomes weaker, while Belle-II becomes stronger, opposite to the situation of smaller ALP masses. These differences in different mass comes from the fact that BR$(a \to \gamma \gamma)$/BR$(a \to e\bar{e})$ is propotional to $m_a^2$. Therefore, for larger $m_a$, the  BR$(a \to \gamma \gamma)$ enhances the product $\sigma$BR and surpass the existing Belle-II limit.
 In contrast, the allowed region for BaBar is increased due to suppression of BR($a\to \bar{e}e$). 

\subsection{Electron beamdump experiments}
ALPs can be searched using electron and positron colliders, especially for masses around $\mathcal{O}(1 \sim 10)$ GeV. However, for lighter ALPs beam dump experiments are cost-effective, since the fixed target or beam dump experiments can provide higher luminosity than colliders. Specifically, beam dump experiments have a low-energy configuration that allows for the effective searches of ALPs with lower masses, ranging from 1 MeV to roughly 1 GeV. In general, electron, muon and proton beams can be used for beamdump searches. In this study, we focus on electron beamdump searches for our purpose of investigating the concurrence scenario of both  $g_{a\bar{e}e}^{\text{eff}}$ and $g_{a\gamma\gamma}^{\text{eff}}$ couplings.

\begin{table}[htb]
  \begin{center}
    \begin{tabular}{|c|c|c|c|c|c|c|}
      \textbf{Experiment} & \textbf{$E_e$[GeV]} & \textbf{Target}&\textbf{$L_{\rm{sh}}$[m]}&\textbf{$L_{\rm{dec}}$[m]}& \textbf{Year}\\
      \hline
      E137~\cite{Bjorken:1988as, Liu:2017htz,Abashian:1980pb} &  20& Al &179&204&1988(SLAC)\\
      NA64(Invis)~\cite{NA64:2020qwq, Dusaev:2020gxi,NA64:2021aiq} & 100& Pb &$\sim$ 4.35&$\infty$&2020(CERN)\\
       KEK~\cite{Ishikawa:2021qna} &  7&W&$\sim$ 0.25&1&2013(KEK linac)\\
            E141~\cite{Riordan:1987aw} &  9.0& W&0.12&35&1987(SLAC)\\
 E774~\cite{Bross:1989mp} &  275& W&0.3&28&1989(Fermilab) \\
Orsay(Higgs)~\cite{Davier:1989wz} &  1.6&W&1&2&1989(LAL)
    \end{tabular}
     \caption{Electron beamdump experiment setups and the relevant parameters for searching ALPs~\cite{Andreas:2012mt}. NA64(Invis) represents the invisible signature configuration of NA64, where ALPs decay beyond all subdetectors of NA64~\cite{Dusaev:2020gxi}.} \label{tab:electronBeamdump}
  \end{center}
\end{table}

In table~\ref{tab:electronBeamdump}, several constraints established through the electron beamdump method have been listed, along with their experimental setups.  Searches such as KEK~\cite{Ishikawa:2021qna}, E141~\cite{Riordan:1987aw}, E774~\cite{Bross:1989mp} and Orsay~\cite{Davier:1989wz} used Tungsten (W) as target materials while the dominant exclusion limits E137~\cite{Bjorken:1988as, Liu:2017htz} and NA64~\cite{NA64:2020qwq, Dusaev:2020gxi,NA64:2021aiq} used Aluminium(Al) and Lead(Pb) respectively. Constraints have been set using E137 on ALP-electron coupling~\cite{Bjorken:1988as, Liu:2017htz} and ALP-photon coupling~\cite{Essig:2010gu, Bjorken:1988as}, and have been set using NA64 on ALP-photon coupling~\cite{NA64:2020qwq, Dusaev:2020gxi} and ALP-electron coupling~\cite{NA64:2021aiq}, but these limits are all considered in single coupling scenario. Therefore, in this study, we recast the ALP limits of E137 and NA64 experiments under the concurrence of the ALP-electron and ALP-photon couplings. 

Both E137 and NA64 constraints consider ALPs produced by incident GeV-electron beam on target, where the ALPs penetrate through shield and decay into electron pairs or photon pairs. The sensitivity of E137 was obtained from the production of pseudoscalar particles where a 20 GeV electron beam was directed at an Aluminium target, penetrating a 179 m long shield and decaying in the 204 m length detector. On the other hand, NA64 experiment utilized a 100 GeV electron beam, passed through one 0.45 m ECAL (Electromagnetic Calorimeter) module, and three 1.3m HCAL (Hadron Calorimeter) modules in the lead target sector. It has two different experimental setups \cite{NA64:2021aiq} and we use the "invisible mode" configuration where the ECAL serves as the target and one HCAL serves as the shield. This configuration still has two search strategies, one for visible signature that ALPs decay in the latter two HCAL modules, while another for invisible signature that the decay position of ALP is assumed to be at the end of the detector (after ECAL and HCAL modules). In our analysis, we consider limits from invisible signature of NA64 and it is in good accordance with the limits which is set under the scenario of single coupling between ALP and electron or photon. 

\begin{figure}[htb]
    \centering
     \begin{subfigure}
    {
        \includegraphics[scale=.75]{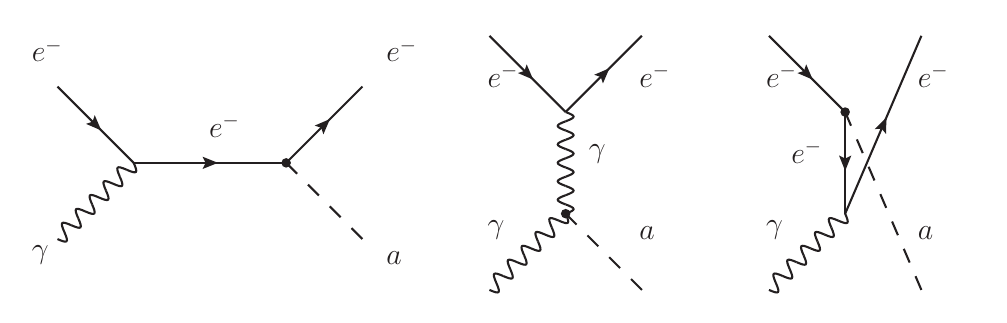}
    }
    \caption{Feynman diagrams for ALP ($a$) associated production in electron beamdump experiments. The left and right diagrams are dark bremsstrahlung process that radiates ALP, while the middle diagram is the Primakoff-like process. We have used IWW approximation to reduce the original 2 to 3 process to a 2 to 2 calculation.
    } \label{fig:beamdumpfeynman}
    \end{subfigure}	
\end{figure}

To estimate the number of detectable ALPs in beam dump experiments, we need to tackle the cross-section of the 2 to 3 production process first: $e^- N\to e^- aN$, where ALP is produced from the collision of incoming electrons with fixed target materials, and this can be done by utilizing the Improved Weizs\"acker-Williams (IWW) approximation method. The concept behind this approach was to simplify the cross section calculation for a 2 to 3 process by reducing it to a 2 to 2 process, simplifying phase space integral. The idea is that the virtual photon emitted from the nuclei can be seemed as real photons under this approximation. This was done to decrease the computational complexity, but it would require mass of the new particle to be significantly smaller than the energy of the incoming beam and result a highly collinear final states. More detailed explanations can be found in refs.~\cite{Tsai:1966js, Kim:1973he, Tsai:1973py, Tsai:1986tx,Bauer:2018onh}. The Feynman diagrams for the simplified 2 to 2 ALP production are depicted in the middle panel of figure~\ref{fig:beamdumpfeynman}. One contribution comes from the Primakoff-like process (the middle panel) via ALP-photon coupling, and there are also contributions from dark bremsstrahlung process (the left and right panels) via ALP-electron coupling.

With IWW approximation, the fraction of cross-section distribution over the energy transfer ratio ($x = E_a/E$) is represented by~\cite{Tsai:1966js,Liu:2017htz}:
\bea\label{eq:dsigmadx}
\frac{d \sigma}{d x} = \frac{\alpha_{\text{QED}}}{4 \pi^2} 
\frac{\sqrt{E^2 x^2 -m^2_a}}{E} \chi \int d \tilde{u} \frac{\mathcal{A}}{\tilde{u}^2} \frac{1-x}{x},
\eea
where $E$ is the energy of initial-state electrons, $\mathcal{A}$ is the approximate 2 to 2 amplitude  and $\chi$ is an effective flux of photons~\cite{Gninenko:2017yus}
\bea
\chi = \int^{m^2_a+m^2_e}_{t_{\text{min}}} \frac{t - t_{\text{min}} }{t} F^2(t),\,\, t_{\text{min}} = \left(\frac{m^2_a}{2 E} \right)^2,
\eea
We have used modified Mandelstam variables $\tilde{s}\equiv s - m^2_e$, $\tilde{u}\equiv u - m^2_e$ and $\tilde{t}\equiv t$~\cite{Liu:2017htz} for convenience. The elastic form factor $F(t)$ is given as~\cite{Bjorken:2009mm,Kim:1973he,Schiff:1951zza}
\bea
F(t) = \left(\frac{a^2 t}{1+a^2t}\right) \left(\frac{1}{1+ t/d}\right)Z ,
\eea
which consists of two components. The first component is the elastic atomic form factor with $a=111\, Z^{1/3}/m_e$, and the second component is the elastic nuclear form factor (nuclear size) with $d=0.164\, \text{GeV}^{-1}\, A^{-2/3}$,
where $Z,A$ are the target atomic number and atomic weight respectively (e.g., Al$^A_Z=$ Al$^{27}_{13}$.). Because of the negligible inelastic form factor contribution, we use the same assumption, which only accounts for the elastic part component as refs.~\cite{Bjorken:2009mm, Liu:2016mqv}. In eq.~(\ref{eq:dsigmadx}), the integrand inside the integral is calculated as:
\bea 
\frac{\mathcal{A}}{\tilde{u}^2}\frac{ 1-x}{ x}&=& \alpha _{\text{QED}} \pi \bigg[ \frac{  (g^{\text{eff}}_{a\gamma\gamma})^2 \left(x^2-2 x+2\right) }{\Delta  x} \nonumber \\&+& \frac{2  m_e^2 x  \bigg(2 \Delta ^2 (g^{\text{eff}}_{a\bar{e}e})^2 x^2
   \left(\Delta ^2+m_a^4 (x-1)^2\right)+2 \Delta  g^{\text{eff}}_{a\bar{e}e} g^{\text{eff}}_{a\gamma\gamma}  \tilde{u}^3  x^5 \bigg)}{\Delta ^2\tilde{u}^4  x^4} \nonumber \\&+& \frac{2  m_e^2 x  \bigg((g^{\text{eff}}_{a\gamma\gamma})^2 (x-1)\tilde{u}^4  x^4\bigg)}{\Delta ^2\tilde{u}^4  x^4}\nonumber\\ &-&\frac{8 m_e^4 \left( ( g^{\text{eff}}_{a\bar{e}e})^2 m_a^2 (x-1) x^5
 \right)}{\tilde{u}^4  x^4}\bigg]+\mathcal{O}\left(m_e^5\right),
\eea
where $\Delta =m^2_a (x-1)-\tilde{u}  x$. Under the condition that $\tilde{t}$ approaches to $t_{\text{min}} \approx m^2_a/(4 E^2)$, these modified Mandelstam variables can be approximated as follows:
\bea
\tilde{s} &=& -\frac{\tilde{u}}{1-x}, \nonumber\\
\tilde{u} &=& - x E^2 \theta^2_{a} - m^2_a \frac{(1-x)}{x} - m^2_e x, \nonumber\\
\tilde{t} &=& \frac{\tilde{u} x}{(1-x)} + m^2_a,
\eea
where $\theta_a$ is the scattering angle between the incoming electron and the produced ALP. We have taken $\theta_a$ to be about $4.4\times 10^{-3}$ for E137 following ref.~\cite{Tsai:1966js}, while for NA64, $\theta_a$ can be extended to infinity because the cross-section is only contributed from small $\theta_a$ ~\cite{Dusaev:2020gxi}.

Finally, the number of detectable ALP can be estimated by the following equation~\cite{Liu:2017htz,Tsai:1966js,Bauer:2018onh} :
\bea \label{eq:ebeamdumpE137}
N_{a} &\approx& \frac{N_e X}{M_{\text{target}}} \int^{E_0}_{E_{\text{min}}} dE \int^{x_{\text{max}}}_{x_{\text{min}}} dx \int^T_0 dt I_e(E_0, E, t) \frac{d\sigma}{d x} e^{- L_{\text{sh}}\left(\frac{1}{l_{a}} + \frac{1}{l_\lambda}\right)} \left(1 - e ^{-\frac{L_{\text{dec}}}{ l_a}}\right) ,
\eea
where $N_e$ is the number of incident electrons, which are $1.87\times 10^{20}$ for E137 and $5\times 10^{12}$ for NA64 in this search. $X$ represents the unit length of the radiation target. $M_{\text{target}}$ is the molar mass of the target atom. $E_0$ is the energy of the initial incident electron beam, which are 20 GeV for E137 and 100 GeV for NA64. The mininal detectable energy is $E_{\text{min}} = m_e + E_{\text{cut}}$, with a typical $E_{\text{cut}}$ value of around 2 GeV for E137, and 20 GeV for NA64. The energy distribution of electrons after passing through a radiation length $t$ is described by the function $I_e$ and we have used the approximation (eq.~(12)) in ref.~\cite{Tsai:1986tx}. The upper bound of $t$ is $T=\rho L_{sh}/X$ where $\rho$ is the density of the target. In the final part of the equation, $L_{\text{sh}}$ and $L_{\text{dec}}$ are the lengths of shield and the distance from the shield to the detector, respectively, which can be found in table~\ref{tab:electronBeamdump}. $l_{\lambda}$ is the absorption of ALP by electrons in the target, which would result in an attenuation factor $\text{exp}(-L_{\text{sh}} /l_{\lambda})$ significant for the thick target~\cite{Tsai:1966js}, but this absorption effect is quite small for our configurations and can be neglected. Finally, the ALP decay length, $l_a$, is related to the parameters $g^{\text{eff}}_{a\bar{e}e}$, $g^{\text{eff}}_{a\gamma\gamma}$, $m_a$ and the energy of the outgoing ALP ($E_a$) which can be defined as follows: 
\bea
l_a = \frac{E_a}{m_a} \frac{1}{\Gamma_a}.
\eea
where $\Gamma_a$ represents the sum of the decay rates of ALP into $\bar{e}e$ and $\gamma\gamma$ as described in eqs.~(\ref{eq:axiontoee}) and (\ref{eq:axiontoAA}). The decay of ALP into electrons can only occur when the mass of the ALP, $m_a$, is greater than $2m_e$. This means that events with $m_a < 2m_e$ will not be considered. The final states of the ALP have been determined to align with the specific coupling restrictions. The exiting experimental exclusion limits are obtained by limiting the number of ALP events with either 100\% diphoton or dielectron decay. Based on the experiment's statistics, the bounds of $N_a = 3$ have been set at a 95\% confidence level for E137, and bounds of $N_a=2.3$ at a 90\% confidence level for NA64, where any results exceeding these values will not be accepted. 

\begin{figure}[t!]
    \centering
     \begin{subfigure}
    {
        \includegraphics[scale=.31]{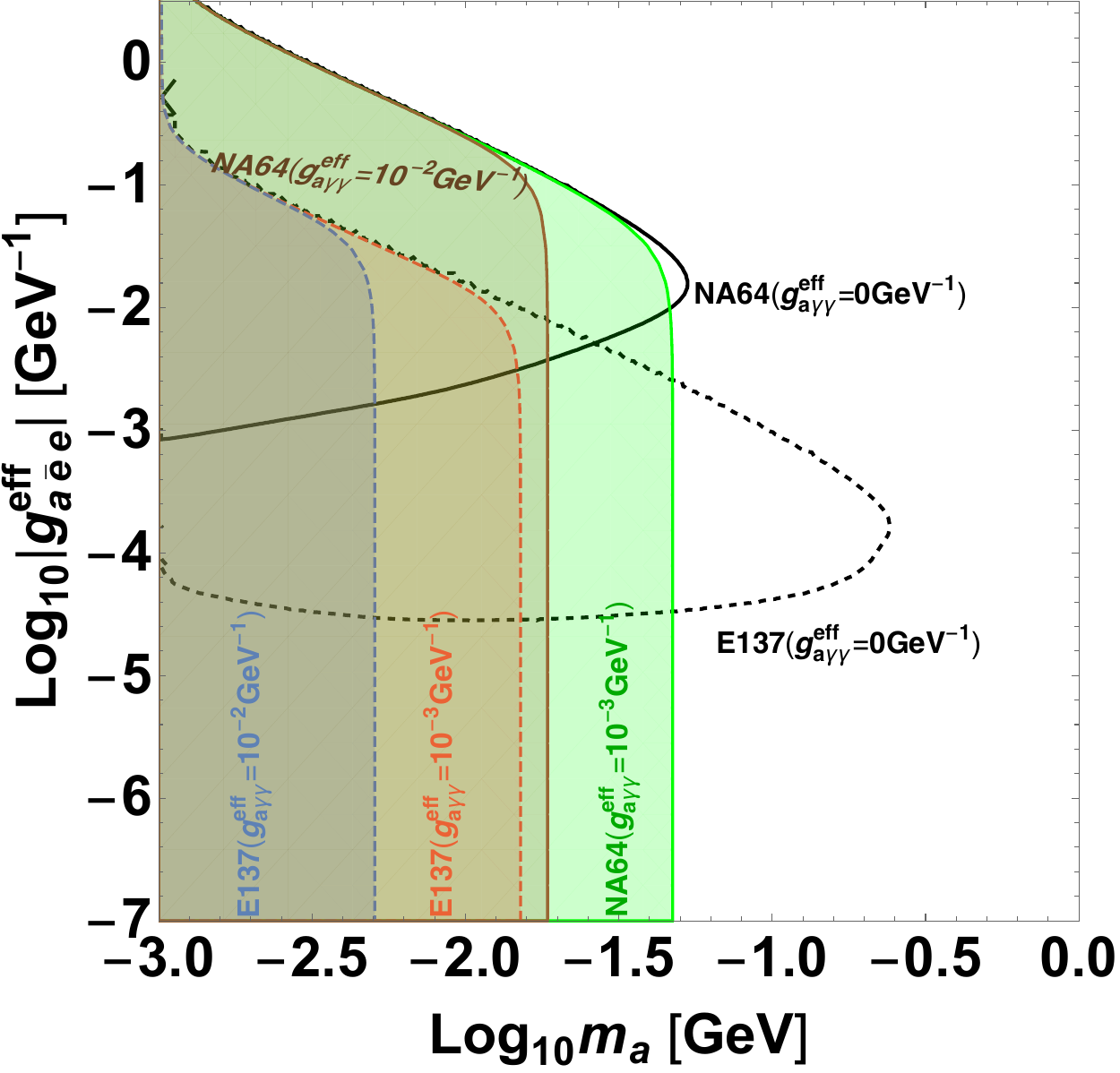}\hspace{1cm}
         \includegraphics[scale=.31]{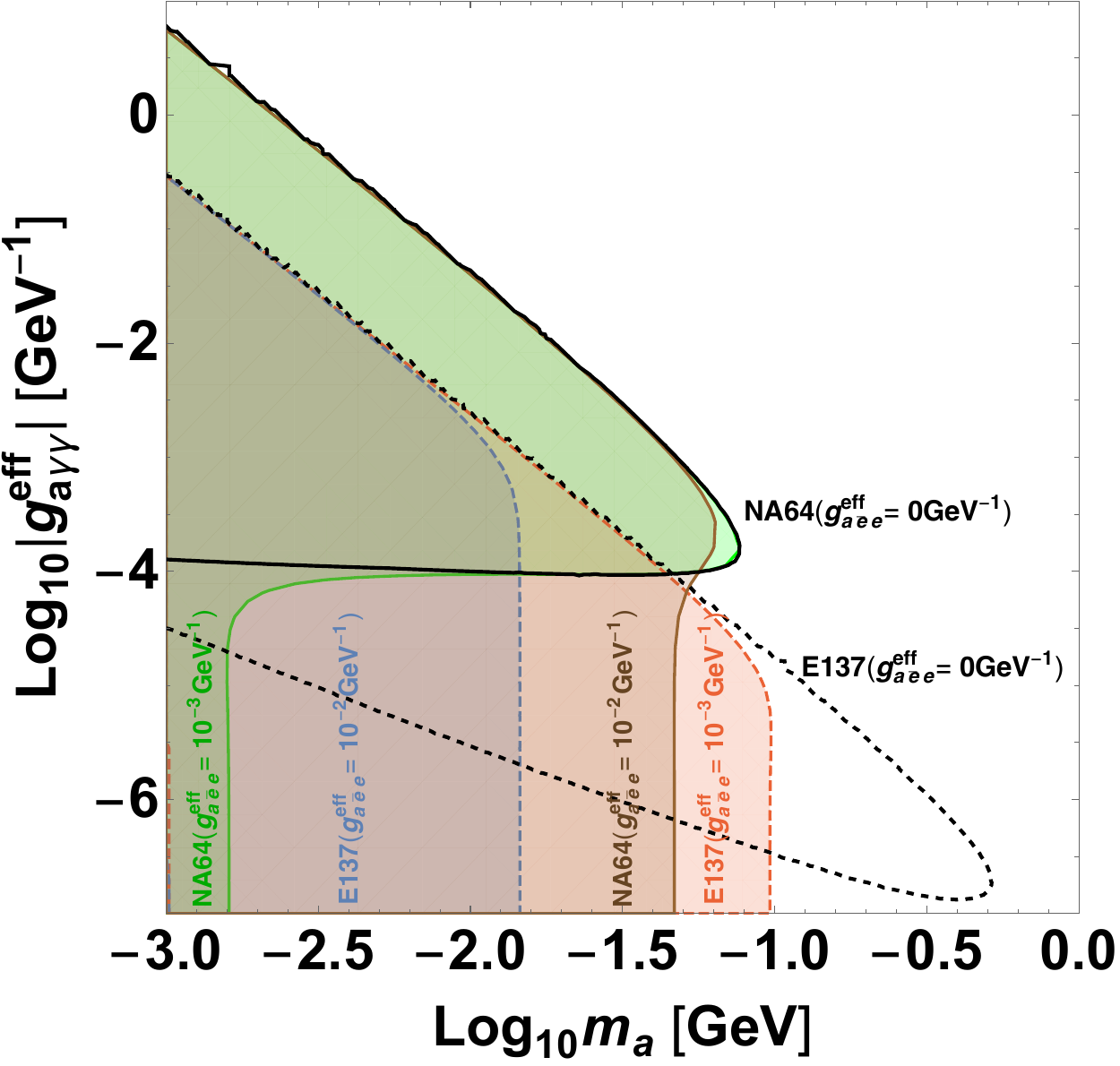}
    }
    \caption{ The existing exclusion region for a simple ALP particle that couples to photons and electrons/positrons is shown by the black solid curve (NA64) and black dashed curve (E137). The filled color areas represent the recasted limits for different coupling choices. In the left panel, the exclusion region for E137 is shown in dashed orange ($|g^{\text{eff}}_{a\gamma\gamma}| = 10^{-3}$  GeV$^{-1}$) and dashed blue ($|g^{\text{eff}}_{a\gamma\gamma}| = 10^{-2}$ GeV$^{-1}$) areas  in the [$ m_a$, $|g^{\text{eff}}_{a\bar{e}e}|$] plane and displays $N_a$. The exclusion limit for recasted NA64, on the other hand, is shown in solid green and solid brown areas, respectively. In the right panel, the exclusion region for recasted E137 and NA64 used same color but fixed $|g^{\text{eff}}_{a\bar{e}e}|$ in the [$ m_a$, $|g^{\text{eff}}_{a\gamma\gamma}|$] plane and displays $N_a $. Both plots are shown in logarithmic scale with base 10.
    }\label{fig:ebeamdumpmagaeE137}
    \end{subfigure}	
\end{figure}

By utilizing the analysis above, we are able to study the impact of the concurrent effect of two couplings through the calculation of number of events production using eq.~(\ref{eq:ebeamdumpE137}). In both the E137 and NA64 searches, it is challenging to determine whether the final states of particles are electrons or photons, as there is no magnetic field near the detector to differentiate between them, and the signal is obtained via calorimeter measurement. Thus, the final states of the ALPs from the beam dump approach cannot be specified, and both $e^+e^-$ and $\gamma \gamma$ decay channels should be taken into account. As a result, the number of events $N_a$ will not depend on the branching ratios but will depend on the ALP's cross-section $\sigma$ and its lifetime. In figure~\ref{fig:ebeamdumpmagaeE137}, we display the existing exclusion region for the E137 experiment (black dashed curve) and the NA64 experiment (black solid curve) in the contour plane [$m_a$, $|g^{\text{eff}}_{a\bar{e}e}|$] (left panel) and [$m_a$, $|g^{\text{eff}}_{a\gamma\gamma}|$] (right panel)\footnote{The units of axes for the contours have been converted to base-10 logarithm scale.}. In each panel, we picked two concurrent coupling benchmark points. On the left panel, the excluded regions with dotted curve bounds are the E137 results with the presence of $|g^{\text{eff}}_{a\gamma\gamma}|$ couplings for $10^{-2}$ GeV$^{-1}$ (blue) and $10^{-3}$ GeV$^{-1}$ (orange) while the regions with solid line curve bounds are the NA64 results with the presence of $|g^{\text{eff}}_{a\gamma\gamma}|$ couplings for $10^{-2}$ GeV$^{-1}$ (brown) and $10^{-3}$ GeV$^{-1}$ (green). The geometry configuration of beam dump experiments, characterized by parameters such as $L_{\rm sh}$ and $L_{\rm dec}$, directly influences the limits they can set on ALP properties. The concurrent effect on the ALP's lifetime is particularly important, as a reduced lifetime leads to a shorter decay length. This feature of concurrent couplings results in a contraction of the excluded region towards lower ALP masses, as decay length is inversely proportional to $m_a$. 
The second feature pertains to the expansion of exclusion regions for low $|g^{\text{eff}}_{a\bar{e}e}|$ couplings in the presence of $|g^{\text{eff}}_{a\gamma\gamma}|$ couplings. In the original scenario with only a single small coupling, fewer events would be produced due to the reduction in cross-section caused by the smaller couplings. As a result, the exclusion regions determined from the recasted E137 and NA64 limits will expand towards lower values of $g^{\text{eff}}_{a\bar{e}e}$ couplings because the presence of additional couplings enlarges the boundaries of the limits since more events would be produced. 

Similarly, in the right panel of figure~\ref{fig:ebeamdumpmagaeE137}, we plot the excluded region under $[m_a,|g^{\text{eff}}_{a\gamma\gamma}|]$ plane for recast E137 solid curve bounds with presence of $|g^{\text{eff}}_{a\bar{e}e}|$ couplings for $10^{-2}$ GeV$^{-1}$ (blue), $10^{-3}$ GeV$^{-1}$ (orange) and recast NA64 dashed curve bounds with presence of $|g^{\text{eff}}_{a\bar{e}e}|$ couplings for $10^{-2}$ GeV$^{-1}$ (brown), $10^{-3}$ GeV$^{-1}$ (green). The regions inside curves are again ruled out with the different selection of $|g^{\text{eff}}_{a\bar{e}e}|$ couplings. Our calculated existing NA64 (E137) limits for ALP with solid black (dashed black) are in good accordance with figure 7 in ref.~\cite{Dusaev:2020gxi}. In general, the presence of two couplings in ALPs leads to shorter lifetimes, which, in turn, causes the exclusion regions to contract towards the regime of small $m_a$. On the other side, the interplay between the two couplings can result in increased events of ALPs in the regime of lower $|g^{\text{eff}}_{a\gamma\gamma}|$ couplings, leading to an enhancement of $N_a$, as mentioned earlier. However, the impact of $|g^{\text{eff}}_{a\gamma\gamma}|$ on the production cross-section is considerably more significant compared to the coupling $|g^{\text{eff}}_{a\bar{e}e}|$. Therefore, the choice of $|g^{\text{eff}}_{a\bar{e}e}|=10^{-3}$ GeV$^{-1}$ is not sufficiently high to yield substantial enhancements, resulting in an exclusion region that is only apparent for ALPs with masses around $m_a \sim 10^{-3}$ GeV but boundaries disappear for heavier ALPs.   

\begin{figure}[htb]
    \centering
     \begin{subfigure}
    {
        \includegraphics[scale=.3]{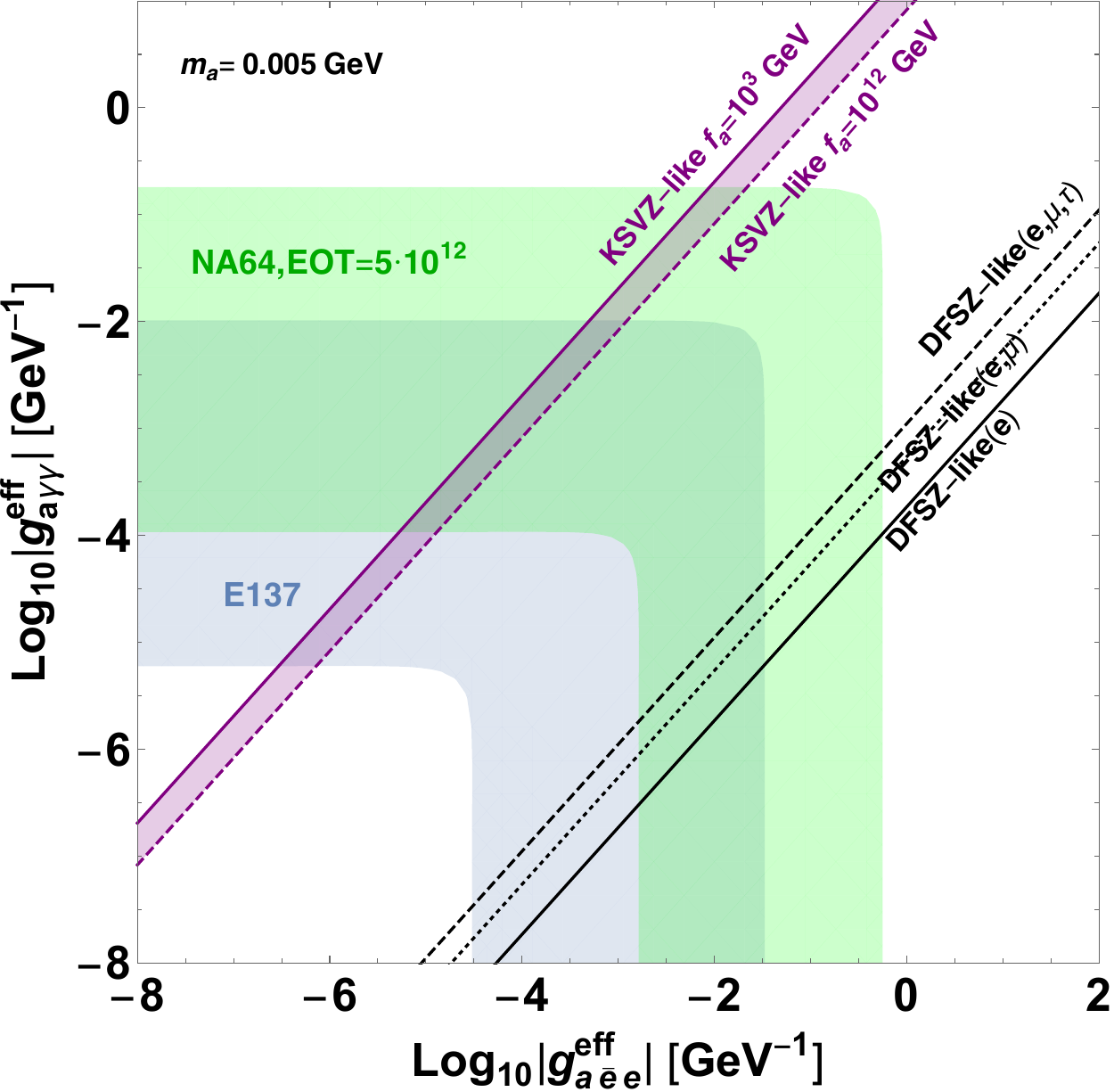}\hspace{1cm}
         \includegraphics[scale=.3]{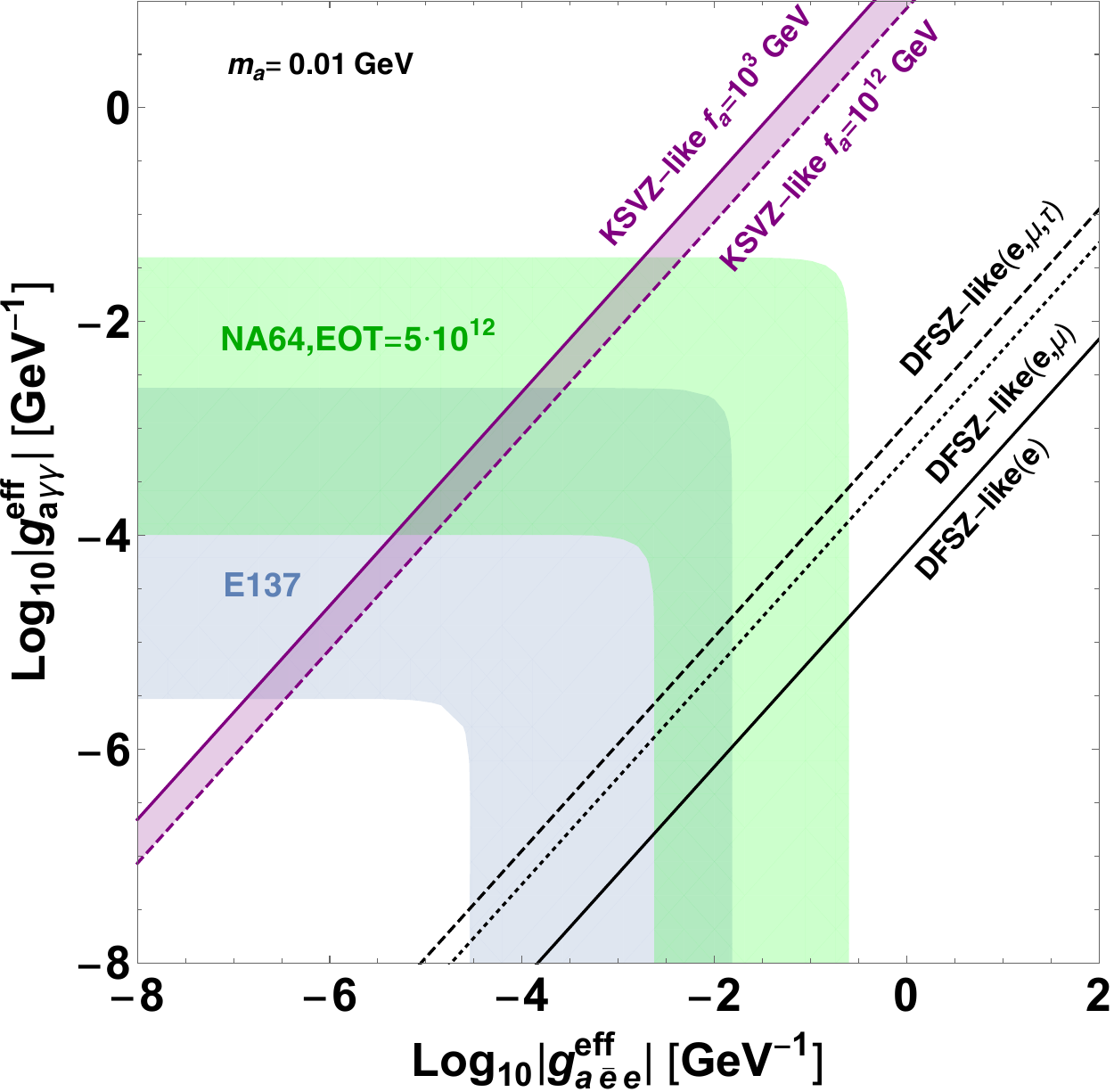}
    }
    \subfigure
   {
         \includegraphics[scale=.3]{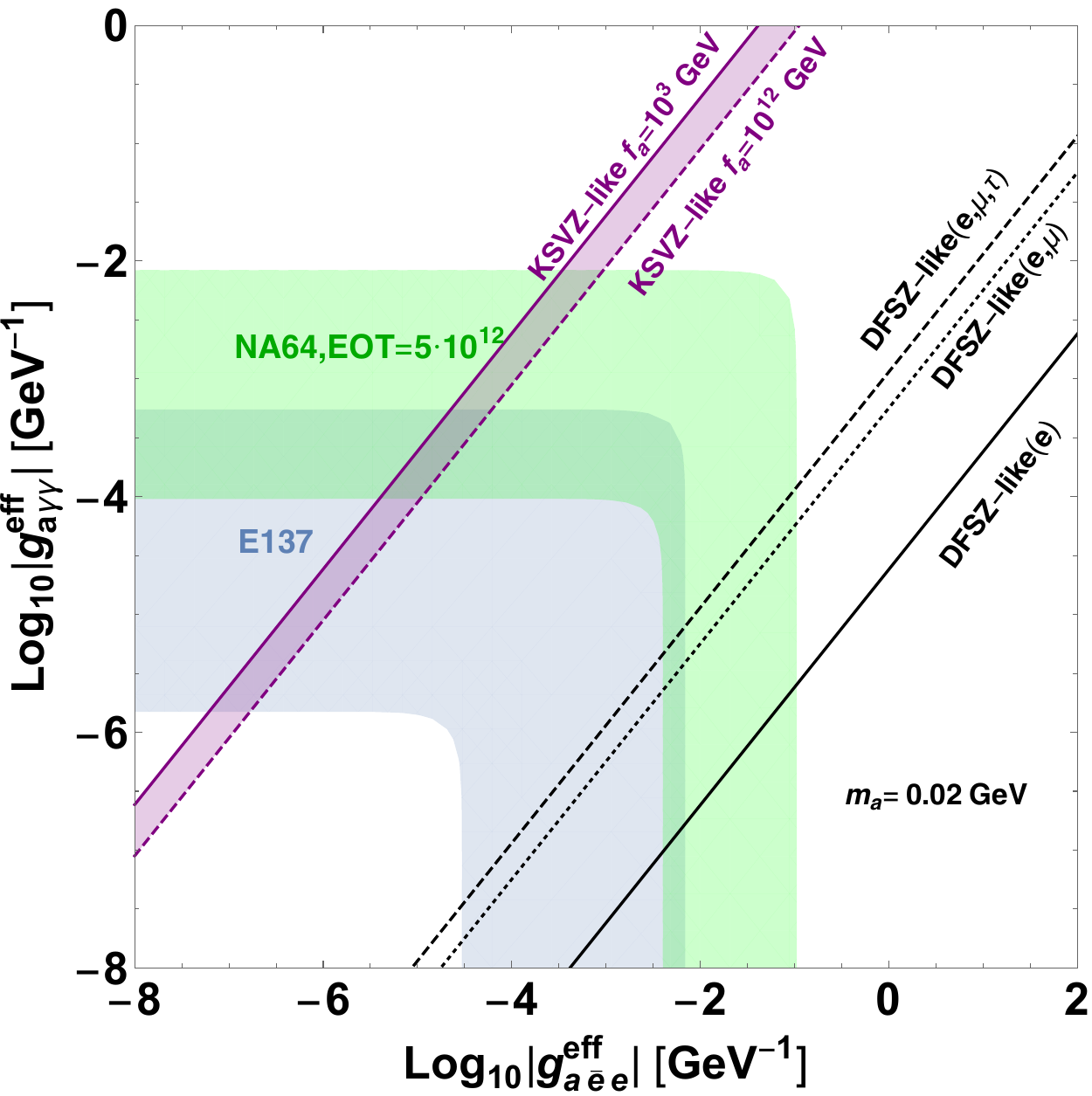}\hspace{1cm}
         \includegraphics[scale=.3]{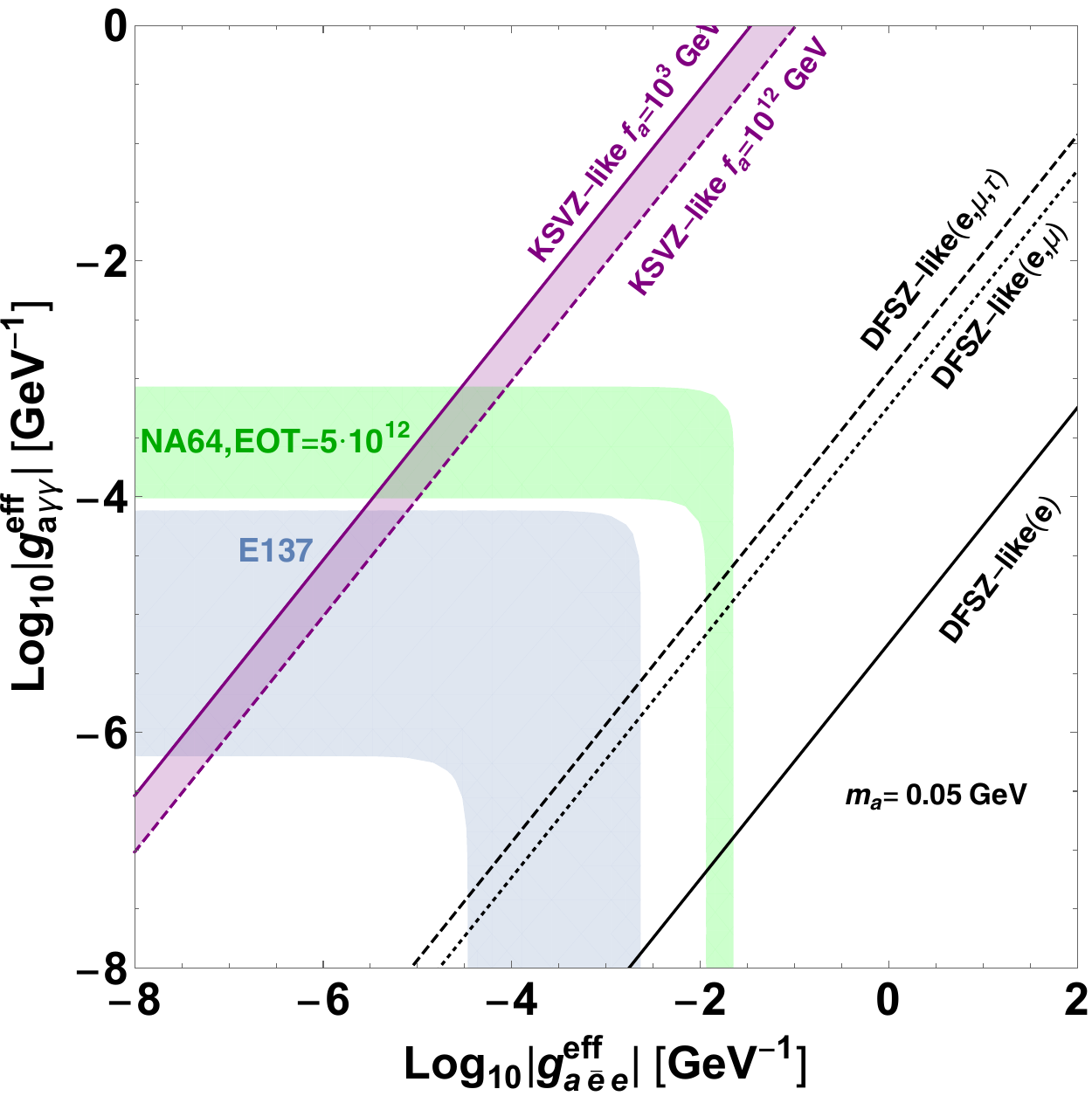}
    }
    \caption{ The excluded region for a simple ALP particle with four different ALP mass choices: $m_a = 0.005, 0.01,0.02, 0.05$ GeV (top left, top right, bottom left, bottom right ) couples to photon ($|g^{\text{eff}}_{a\gamma\gamma}|$) and electron/positron ($|g^{\text{eff}}_{a\bar{e}e}|$) in the [$|g^{\text{eff}}_{a\bar{e}e}|, |g^{\text{eff}}_{a\gamma\gamma}|$] plane from E137 ($N_a $ in light blue color) and NA64 ($N_a $ in green color) calculation respectively.  The white color areas are allowed. The black diagonal lines are the DFSZ-like UV model (photon coupling produced from $e$ loop only (solid), $e,\mu$ loops (dotted) and all three leptons $e,\mu, \tau$ loops (dashed)) available region. The purple diagonal parts are allowed the KSVZ-like model with $f_a=10^3$ GeV (solid) and $10^{12}$ GeV(dashed).
    }\label{fig:gaegaA4subplotsbeamdumpE137}
    \end{subfigure}	
\end{figure}

To emphasize the concurrent effects in electron beamdump results based on presence of two couplings, we plotted the constrained parameter space for E137 and NA64 in 2D plane $[|g^{\text{eff}}_{a\bar{e}e}|, |g^{\text{eff}}_{a\gamma\gamma}|]$ with four light ALP mass benchmark points ($m_a <0.1$ GeV) in figure~\ref{fig:gaegaA4subplotsbeamdumpE137}. The green regions represent the NA64 exclusion by equating the number of ALP events ($N_a$) to 2.3 (90\% C.L.). The E137 setup excludes the blue regions by equating the number to 3 (95\% C.L.). The regions outside curves are allowed with $m_a =$ 0.005 GeV (top left panel), 0.01 GeV (top right panel), 0.02 GeV (bottom left panel), and 0.05 GeV (bottom right panel). The mixing of two couplings provides a larger survival region. For example, the existing single $|g^{\text{eff}}_{a\bar{e}e}|$ coupling constraint has been completely excluded in the range of $10^{-4.5}$ to $10^{0}$ GeV$^{-1}$ for $m_a = 0.005$ GeV in single coupling scenario. However, in the top left panel of figure~\ref{fig:gaegaA4subplotsbeamdumpE137}, this constraint can be evaded if $|g^{\text{eff}}_{a\gamma\gamma}| > 10^{-0.5}$ GeV$^{-1}$.  
In figure~\ref{fig:gaegaA4subplotsbeamdumpE137}, we also see the excluded region becomes smaller with an increasing $m_a$.

In addition, we present the UV model preferred parameter spaces in figure~\ref{fig:gaegaA4subplotsbeamdumpE137} to compare with the bean dump limits. 
The UV model preferred region, which survives existing constraints can be classified into two categories. The first category is that the concurrence of both two couplings could relax large coupling regions. Taking $m_a = 0.005$ GeV as an example, in a single coupling scenario, the parameter space about $10^{-5} < |g^{\text{eff}}_{a\gamma\gamma}| < 10^{-0.5}$ GeV$^{-1}$ or $10^{-4.5} < |g^{\text{eff}}_{a\bar{e}e} |< 10^{0}$ GeV$^{-1}$ are fully excluded separately. However, the concurrence of both couplings makes the parameter space open when both couplings are large. The reason is that the number of events $N_a$ is affected by the lifetime of the ALP as before. When $|g^{\text{eff}}_{a\bar{e}e}|$ increases, the decay length of ALP ($l_a$) will decrease, causing a decrease in the $e^{-L_{\text{sh}}/l_a}$ term and thus decrease in $N_a$. The same applies to the relationship between $|g^{\text{eff}}_{a\gamma\gamma}|$ and $l_a$. Therefore, in the concurrence scenario, the upper-right region of the parameter space is opened. There are some modifications in the cross-section, but generally, it can not compete with the effects of lifetime decrease. 

For the UV models, the constraints on both KSVZ-like and DFSZ-like models are significantly relaxed, as shown in figure~\ref{fig:gaegaA4subplotsbeamdumpE137}. The second category of parameter space is the left bottom corner in figure~\ref{fig:gaegaA4subplotsbeamdumpE137} for all $m_a$ (e.g., $|g^{\text{eff}}_{a\bar{e}e}| < 10^{-4.5} $ GeV$^{-1}$ and $|g^{\text{eff}}_{a\gamma\gamma}| < 10^{-6}$ GeV$^{-1}$ in $m_a=0.02$ GeV plane.). The statistical bounds are satisfied when the coupling values are small enough in these corners, leading to a small $N_a$. As $m_a$ increases, one can see that the excluded regions shifted to the left-bottom corner because a mild increase of $m_a$ leads to an increase of $N_a$. One important factor is that the physical decay volume is fixed, which requires a fixed decay length $l_a$. When $m_a$ becomes large, $l_a$ becomes small. To compensate for this change, the couplings should be smaller to keep $l_a$ unchanged. When couplings are changed, the cross-sections are modified as well, but their effect is quadratic. Therefore, the shift is a complex combination of the change in cross-sections and lifetime.

 \begin{figure}[t!]
   	\centering
   	\includegraphics[width=0.6\linewidth]{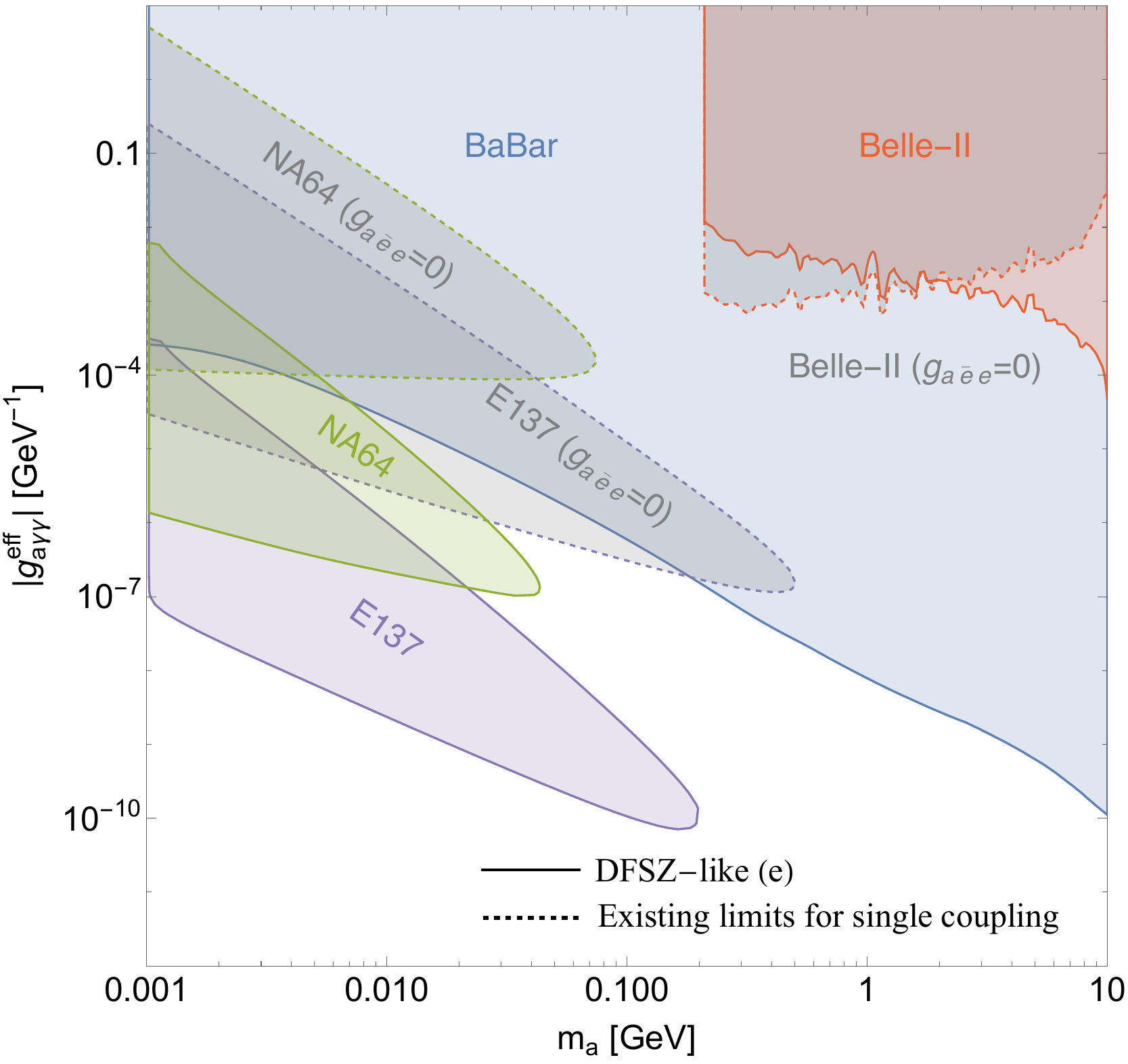} 
   	\caption{The exclusion region depicted in the [$m_a$, $|g^{\text{eff}}_{a\gamma\gamma}|$] parameter space contains constraints from various experiments. Specifically, the constraints from collider experiments BaBar (blue) and Belle-II (red), as well as beam dump experiments E137 (purple) and NA64 (green), are represented. The dashed lines on the plot represent constraints obtained from a single $g^{\text{eff}}_{a\gamma\gamma}$ coupling (with $g^{\text{eff}}_{a\bar{e}e}=0$), while the solid line corresponds to our DFSZ-like (e) model, where $g^{\text{eff}}_{a\gamma\gamma}$ and $g^{\text{eff}}_{a\bar{e}e}$ are correlated. The BaBar limits are indirectly derived from the constraints of $g^{\text{eff}}_{a\bar{e} e}$ and are only applicable within this DFSZ-like (e) model. In such a scenario, we convert the existing constraint of ALP-electron pairs, $g^{\text{eff}}_{a\bar{e}e}$, and translate it into the coupling of ALP-photon pairs, $g^{\text{eff}}_{a\gamma\gamma}$, based on the ratio relationship labeled as eq.~(\ref{eq:caAAovercaee}). It should be noted that in the KSVZ-like model, where $g^{\text{eff}}_{a\gamma\gamma}$ is significantly larger than $g^{\text{eff}}_{a\bar{e}e}$, the modification of these limits is not observable.}
   	\label{fig:dfsz_limits}
   \end{figure}

Furthermore, in order to observe the concurrent effect on previous specific UV model, we present the constraints derived from both collider experiments and beam dump experiments using the DFSZ-like (e) model.  These constraints are plotted in the [$m_a$, $|g^{\text{eff}}_{a\gamma\gamma}|$] plane in figure \ref{fig:dfsz_limits}. We also present the original limits with single $g^{\text{eff}}_{a\gamma\gamma}$ coupling, and the concurrent effect is clearly evident. The downward shift towards smaller values of $g^{\text{eff}}_{a\gamma\gamma}$ for the beam dump constraints can be attributed to the increased production cross-section and reduced lifetime, as previously mentioned. The BaBar limits for $g^{\text{eff}}_{a\gamma\gamma}$ are indirect constraints, applicable only in the context of concurrent UV models. In this scenario, we convert the existing constraint of ALP-electron pairs, $g^{\text{eff}}_{a\bar{e}e}$, to the coupling of ALP-photon pairs, $g^{\text{eff}}_{a\gamma\gamma}$, utilizing the ratio relationship labeled as eq.~(\ref{eq:caAAovercaee}). 
The DFSZ-like (e) model uncovers a small $|g^{\text{eff}}_{a\gamma\gamma}|/|g^{\text{eff}}_{a\bar{e} e}|$ ratio, allowing the translated BaBar limits to extend into the lower ALP-photon coupling regime, mirroring the characteristic displayed in Figure~\ref{fig:UVratiowithMa}. On the other hand, since $g^{\text{eff}}_{a\gamma\gamma}$ is much larger than $g^{\text{eff}}_{a\bar{e}e}$ in the KSVZ-like model, the modification of these limits is not readily observable in this particular plane.

In summary, the concurrent effects of both couplings result in a significant enhancement of the ALP decay width, leading to a reduction of its lifetime and an increase in its production cross-section. This effect is clearly seen in the large coupling regions in figure~\ref{fig:gaegaA4subplotsbeamdumpE137}, where the event rate is significantly reduced. However, in beam dump experiments, the opening of two decay channels is not an issue since both the displaced dielectron and diphoton final states are considered as a signal. Therefore, the branching ratio is less affected compared to electron-positron collider searches. We would like to emphasize that while previous ALP studies have considered the combination bounds between ALP-photon and other couplings, such as $g_{a\bar{e}e}g_{a\gamma\gamma}$ as discussed in~\cite{Xiao:2022rxk}, $g^{\text{eff}}_{aN}g_{a\gamma}$ in~\cite{DiLuzio:2021qct}, $g^{\text{eff}}_{aN}g_{ae}$ in~\cite{EDELWEISS:2018tde} and more, our current work does not specifically address these bounds. The reason for this is that previous studies primarily focused on the stellar and astrophysics constraints, which are applicable to ALPs with masses lighter than the eV scale. In contrast, our study is centered around the MeV to GeV energy range, where beamdump and collider searches provide the dominant limits and have not been covered in any experiments. However, we acknowledge that the combination of these bounds could play a crucial role in future experiments that are capable of exploring the simultaneous effects of multiple couplings. Therefore, it is worth considering the interplay between different coupling bounds when designing future experiments. In the end, our investigation showed that the concurrent scenario is particularly significant for ALP masses in the MeV to GeV range, especially in beam dump searches. The parameter space for ALP survival has two regions, one where both couplings are very small and the other where one of the couplings is very large.

\section{Conclusions}\label{sec: conclu}

In this study, we investigate ALP searches using both photon and electron couplings at electron-positron collider and electron beam dump experiments. This analysis differs from previous research, which determined exclusion limits based on a single coupling without considering the concurrent effect. We have taken two UV models as examples: the KSVZ-like model and the DFSZ-like model. These two UV models predominantly have only one coupling, either the ALP-electron or the ALP-photon coupling. However, the other coupling can be generated from the loop. Therefore, these two couplings are correlated in the UV models, and it is necessary to analyze the ALP limits in the presence of the concurrence of both couplings.

The concurrence of couplings has three physical effects. Firstly, the cross-sections are altered as more diagrams need to be included in the calculation. Secondly, the ALP decay branching ratios change from $100\%$ in the single coupling scenario to two branching ratios, one for dielectron and the other for diphoton decay. Thirdly, the ALP decay width is enhanced due to the opening of a new decay channel, which reduces the lifetime of ALP.
In the electron-positron collider experiment, both UV models have limited options to satisfy the constraints. Our analysis indicates that the limits on the KSVZ-like model are logarithmically dependent on the energy scale and mass of ALPs, while the limits on DFSZ-like models only vary for ($e,\mu$) and ($e,\mu,\tau$) couplings. The interference in the cross-section is minimal when both couplings are considered. Consequently, while the cross-sections increase due to the presence of the additional coupling, the effect is counteracted by the decrease in the relevant branching ratio.

In beam dump experiments, the concurrent effects have a more significant impact on the limits. This is mainly due to the reduction in the lifetime of ALP, which shifts the exclusion region towards lower masses. However, the concurrence also increases the production cross-section which could exclude the region where there is one coupling small but the other coupling is large, because the beam dump experiments can not distinguish the two decay final states. 
Another factor is the presence of large coupling regions that survive due to the fast decay of ALP, leading to a decrease in the number of events in beam dump experiments. Both UV models exhibit significant opening spaces in the small coupling and large coupling regions. Additionally, the shape of the exclusion regions is altered compared to the single coupling scenario. Consequently, the low mass regions of ALP are less constrained by beam dump experiments in the concurrence scenario comparing with the single coupling scenario.

\acknowledgments
The work of JL is supported by NSFC under Grant No. 12075005, 12235001 and by Peking University under startup Grant No. 7101502458.

\bibliographystyle{JHEP}
\bibliography{refe}

\providecommand{\href}[2]{#2}\begingroup\raggedright\begin{thebibliography}{10}

\bibitem{Peccei:1977hh}
R.D.~Peccei and H.R.~Quinn, \emph{{CP Conservation in the Presence of
  Instantons}}, \href{https://doi.org/10.1103/PhysRevLett.38.1440}{\emph{Phys.
  Rev. Lett.} {\bfseries 38} (1977) 1440}.

\bibitem{Peccei:1977ur}
R.D.~Peccei and H.R.~Quinn, \emph{{Constraints Imposed by CP Conservation in
  the Presence of Instantons}},
  \href{https://doi.org/10.1103/PhysRevD.16.1791}{\emph{Phys. Rev. D}
  {\bfseries 16} (1977) 1791}.

\bibitem{Weinberg:1977ma}
S.~Weinberg, \emph{{A New Light Boson?}},
  \href{https://doi.org/10.1103/PhysRevLett.40.223}{\emph{Phys. Rev. Lett.}
  {\bfseries 40} (1978) 223}.

\bibitem{Wilczek:1977pj}
F.~Wilczek, \emph{{Problem of Strong $P$ and $T$ Invariance in the Presence of
  Instantons}}, \href{https://doi.org/10.1103/PhysRevLett.40.279}{\emph{Phys.
  Rev. Lett.} {\bfseries 40} (1978) 279}.

\bibitem{Kim:1979if}
J.E.~Kim, \emph{{Weak Interaction Singlet and Strong CP Invariance}},
  \href{https://doi.org/10.1103/PhysRevLett.43.103}{\emph{Phys. Rev. Lett.}
  {\bfseries 43} (1979) 103}.

\bibitem{Peccei:2006as}
R.D.~Peccei, \emph{{The Strong CP problem and axions}},
  \href{https://doi.org/10.1007/978-3-540-73518-2_1}{\emph{Lect. Notes Phys.}
  {\bfseries 741} (2008) 3}
  [\href{https://arxiv.org/abs/hep-ph/0607268}{{\ttfamily hep-ph/0607268}}].

\bibitem{Arvanitaki:2009fg}
A.~Arvanitaki, S.~Dimopoulos, S.~Dubovsky, N.~Kaloper and J.~March-Russell,
  \emph{{String Axiverse}},
  \href{https://doi.org/10.1103/PhysRevD.81.123530}{\emph{Phys. Rev. D}
  {\bfseries 81} (2010) 123530}
  [\href{https://arxiv.org/abs/0905.4720}{{\ttfamily 0905.4720}}].

\bibitem{Raffelt:1990yz}
G.G.~Raffelt, \emph{{Astrophysical methods to constrain axions and other novel
  particle phenomena}},
  \href{https://doi.org/10.1016/0370-1573(90)90054-6}{\emph{Phys. Rept.}
  {\bfseries 198} (1990) 1}.

\bibitem{Sloan:2016aub}
J.V.~Sloan et~al., \emph{{Limits on axion\textendash{}photon coupling or on
  local axion density: Dependence on models of the Milky Way\textquoteright{}s
  dark halo}}, \href{https://doi.org/10.1016/j.dark.2016.09.003}{\emph{Phys.
  Dark Univ.} {\bfseries 14} (2016) 95}.

\bibitem{Raffelt:2006cw}
G.G.~Raffelt, \emph{{Astrophysical axion bounds}},
  \href{https://doi.org/10.1007/978-3-540-73518-2_3}{\emph{Lect. Notes Phys.}
  {\bfseries 741} (2008) 51}
  [\href{https://arxiv.org/abs/hep-ph/0611350}{{\ttfamily hep-ph/0611350}}].

\bibitem{Asztalos:2006kz}
S.J.~Asztalos, L.J.~Rosenberg, K.~van Bibber, P.~Sikivie and K.~Zioutas,
  \emph{{Searches for astrophysical and cosmological axions}},
  \href{https://doi.org/10.1146/annurev.nucl.56.080805.140513}{\emph{Ann. Rev.
  Nucl. Part. Sci.} {\bfseries 56} (2006) 293}.

\bibitem{Kawasaki:2013ae}
M.~Kawasaki and K.~Nakayama, \emph{{Axions: Theory and Cosmological Role}},
  \href{https://doi.org/10.1146/annurev-nucl-102212-170536}{\emph{Ann. Rev.
  Nucl. Part. Sci.} {\bfseries 63} (2013) 69}
  [\href{https://arxiv.org/abs/1301.1123}{{\ttfamily 1301.1123}}].

\bibitem{Dafni:2018tvj}
T.~Dafni, C.A.J.~O'Hare, B.~Laki\'c, J.~Gal\'an, F.J.~Iguaz, I.G.~Irastorza
  et~al., \emph{{Weighing the solar axion}},
  \href{https://doi.org/10.1103/PhysRevD.99.035037}{\emph{Phys. Rev. D}
  {\bfseries 99} (2019) 035037}
  [\href{https://arxiv.org/abs/1811.09290}{{\ttfamily 1811.09290}}].

\bibitem{AxionLimits}
C.~O'Hare, ``cajohare/axionlimits: Axionlimits.''
  \url{https://cajohare.github.io/AxionLimits/}, July, 2020.
\newblock 10.5281/zenodo.3932430.

\bibitem{Dobrich:2015jyk}
B.~D\"obrich, J.~Jaeckel, F.~Kahlhoefer, A.~Ringwald and K.~Schmidt-Hoberg,
  \emph{{ALPtraum: ALP production in proton beam dump experiments}},
  \href{https://doi.org/10.1007/JHEP02(2016)018}{\emph{JHEP} {\bfseries 02}
  (2016) 018} [\href{https://arxiv.org/abs/1512.03069}{{\ttfamily
  1512.03069}}].

\bibitem{Bjorken:1988as}
J.D.~Bjorken, S.~Ecklund, W.R.~Nelson, A.~Abashian, C.~Church, B.~Lu et~al.,
  \emph{{Search for Neutral Metastable Penetrating Particles Produced in the
  SLAC Beam Dump}}, \href{https://doi.org/10.1103/PhysRevD.38.3375}{\emph{Phys.
  Rev. D} {\bfseries 38} (1988) 3375}.

\bibitem{NA64:2020qwq}
{\scshape NA64} collaboration, \emph{{Search for Axionlike and Scalar Particles
  with the NA64 Experiment}},
  \href{https://doi.org/10.1103/PhysRevLett.125.081801}{\emph{Phys. Rev. Lett.}
  {\bfseries 125} (2020) 081801}
  [\href{https://arxiv.org/abs/2005.02710}{{\ttfamily 2005.02710}}].

\bibitem{Dusaev:2020gxi}
R.R.~Dusaev, D.V.~Kirpichnikov and M.M.~Kirsanov, \emph{{Photoproduction of
  axionlike particles in the NA64 experiment}},
  \href{https://doi.org/10.1103/PhysRevD.102.055018}{\emph{Phys. Rev. D}
  {\bfseries 102} (2020) 055018}
  [\href{https://arxiv.org/abs/2004.04469}{{\ttfamily 2004.04469}}].

\bibitem{NA64:2021aiq}
{\scshape NA64} collaboration, \emph{{Search for pseudoscalar bosons decaying
  into $e^+e^-$ pairs in the NA64 experiment at the CERN SPS}},
  \href{https://doi.org/10.1103/PhysRevD.104.L111102}{\emph{Phys. Rev. D}
  {\bfseries 104} (2021) L111102}
  [\href{https://arxiv.org/abs/2104.13342}{{\ttfamily 2104.13342}}].

\bibitem{Kleban:2005rj}
M.~Kleban and R.~Rabadan, \emph{{Collider bounds on pseudoscalars coupling to
  gauge bosons}},  \href{https://arxiv.org/abs/hep-ph/0510183}{{\ttfamily
  hep-ph/0510183}}.

\bibitem{Mimasu:2014nea}
K.~Mimasu and V.~Sanz, \emph{{ALPs at Colliders}},
  \href{https://doi.org/10.1007/JHEP06(2015)173}{\emph{JHEP} {\bfseries 06}
  (2015) 173} [\href{https://arxiv.org/abs/1409.4792}{{\ttfamily 1409.4792}}].

\bibitem{Brivio:2017ije}
I.~Brivio, M.B.~Gavela, L.~Merlo, K.~Mimasu, J.M.~No, R.~del Rey et~al.,
  \emph{{ALPs Effective Field Theory and Collider Signatures}},
  \href{https://doi.org/10.1140/epjc/s10052-017-5111-3}{\emph{Eur. Phys. J. C}
  {\bfseries 77} (2017) 572}
  [\href{https://arxiv.org/abs/1701.05379}{{\ttfamily 1701.05379}}].

\bibitem{Batell:2009yf}
B.~Batell, M.~Pospelov and A.~Ritz, \emph{{Probing a Secluded U(1) at
  B-factories}}, \href{https://doi.org/10.1103/PhysRevD.79.115008}{\emph{Phys.
  Rev. D} {\bfseries 79} (2009) 115008}
  [\href{https://arxiv.org/abs/0903.0363}{{\ttfamily 0903.0363}}].

\bibitem{Belle-II:2020jti}
{\scshape Belle-II} collaboration, \emph{{Search for Axion-Like Particles
  produced in $e^+e^-$ collisions at Belle II}},
  \href{https://doi.org/10.1103/PhysRevLett.125.161806}{\emph{Phys. Rev. Lett.}
  {\bfseries 125} (2020) 161806}
  [\href{https://arxiv.org/abs/2007.13071}{{\ttfamily 2007.13071}}].

\bibitem{BaBar:2014zli}
{\scshape BaBar} collaboration, \emph{{Search for a Dark Photon in $e^+e^-$
  Collisions at BaBar}},
  \href{https://doi.org/10.1103/PhysRevLett.113.201801}{\emph{Phys. Rev. Lett.}
  {\bfseries 113} (2014) 201801}
  [\href{https://arxiv.org/abs/1406.2980}{{\ttfamily 1406.2980}}].

\bibitem{Bauer:2017ris}
M.~Bauer, M.~Neubert and A.~Thamm, \emph{{Collider Probes of Axion-Like
  Particles}}, \href{https://doi.org/10.1007/JHEP12(2017)044}{\emph{JHEP}
  {\bfseries 12} (2017) 044}
  [\href{https://arxiv.org/abs/1708.00443}{{\ttfamily 1708.00443}}].

\bibitem{Bauer:2018uxu}
M.~Bauer, M.~Heiles, M.~Neubert and A.~Thamm, \emph{{Axion-Like Particles at
  Future Colliders}},
  \href{https://doi.org/10.1140/epjc/s10052-019-6587-9}{\emph{Eur. Phys. J. C}
  {\bfseries 79} (2019) 74} [\href{https://arxiv.org/abs/1808.10323}{{\ttfamily
  1808.10323}}].

\bibitem{Bauer:2020jbp}
M.~Bauer, M.~Neubert, S.~Renner, M.~Schnubel and A.~Thamm, \emph{{The
  Low-Energy Effective Theory of Axions and ALPs}},
  \href{https://doi.org/10.1007/JHEP04(2021)063}{\emph{JHEP} {\bfseries 04}
  (2021) 063} [\href{https://arxiv.org/abs/2012.12272}{{\ttfamily
  2012.12272}}].

\bibitem{Bauer:2021mvw}
M.~Bauer, M.~Neubert, S.~Renner, M.~Schnubel and A.~Thamm, \emph{{Flavor probes
  of axion-like particles}},
  \href{https://doi.org/10.1007/JHEP09(2022)056}{\emph{JHEP} {\bfseries 09}
  (2022) 056} [\href{https://arxiv.org/abs/2110.10698}{{\ttfamily
  2110.10698}}].

\bibitem{Arias-Aragon:2022iwl}
F.~Arias-Arag\'on, J.~Quevillon and C.~Smith, \emph{{Axion-like ALPs}},
  \href{https://doi.org/10.1007/JHEP03(2023)134}{\emph{JHEP} {\bfseries 03}
  (2023) 134} [\href{https://arxiv.org/abs/2211.04489}{{\ttfamily
  2211.04489}}].

\bibitem{Xiao:2022rxk}
M.~Xiao, P.~Carenza, M.~Giannotti, A.~Mirizzi, K.M.~Perez, O.~Straniero et~al.,
  \emph{{Betelgeuse constraints on coupling between axionlike particles and
  electrons}}, \href{https://doi.org/10.1103/PhysRevD.106.123019}{\emph{Phys.
  Rev. D} {\bfseries 106} (2022) 123019}
  [\href{https://arxiv.org/abs/2204.03121}{{\ttfamily 2204.03121}}].

\bibitem{Gao:2020wer}
C.~Gao, J.~Liu, L.-T.~Wang, X.-P.~Wang, W.~Xue and Y.-M.~Zhong,
  \emph{{Reexamining the Solar Axion Explanation for the XENON1T Excess}},
  \href{https://doi.org/10.1103/PhysRevLett.125.131806}{\emph{Phys. Rev. Lett.}
  {\bfseries 125} (2020) 131806}
  [\href{https://arxiv.org/abs/2006.14598}{{\ttfamily 2006.14598}}].

\bibitem{DiLuzio:2021qct}
L.~Di~Luzio et~al., \emph{{Probing the axion\textendash{}nucleon coupling with
  the next generation of~axion helioscopes}},
  \href{https://doi.org/10.1140/epjc/s10052-022-10061-1}{\emph{Eur. Phys. J. C}
  {\bfseries 82} (2022) 120}
  [\href{https://arxiv.org/abs/2111.06407}{{\ttfamily 2111.06407}}].

\bibitem{Bonilla:2021ufe}
J.~Bonilla, I.~Brivio, M.B.~Gavela and V.~Sanz, \emph{{One-loop corrections to
  ALP couplings}}, \href{https://doi.org/10.1007/JHEP11(2021)168}{\emph{JHEP}
  {\bfseries 11} (2021) 168}
  [\href{https://arxiv.org/abs/2107.11392}{{\ttfamily 2107.11392}}].

\bibitem{Alonso-Alvarez:2018irt}
G.~Alonso-\'Alvarez, M.B.~Gavela and P.~Quilez, \emph{{Axion couplings to
  electroweak gauge bosons}},
  \href{https://doi.org/10.1140/epjc/s10052-019-6732-5}{\emph{Eur. Phys. J. C}
  {\bfseries 79} (2019) 223}
  [\href{https://arxiv.org/abs/1811.05466}{{\ttfamily 1811.05466}}].

\bibitem{Ertas:2020xcc}
F.~Ertas and F.~Kahlhoefer, \emph{{On the interplay between astrophysical and
  laboratory probes of MeV-scale axion-like particles}},
  \href{https://doi.org/10.1007/JHEP07(2020)050}{\emph{JHEP} {\bfseries 07}
  (2020) 050} [\href{https://arxiv.org/abs/2004.01193}{{\ttfamily
  2004.01193}}].

\bibitem{Darme:2020sjf}
L.~Darm\'e, F.~Giacchino, E.~Nardi and M.~Raggi, \emph{{Invisible decays of
  axion-like particles: constraints and prospects}},
  \href{https://doi.org/10.1007/JHEP06(2021)009}{\emph{JHEP} {\bfseries 06}
  (2021) 009} [\href{https://arxiv.org/abs/2012.07894}{{\ttfamily
  2012.07894}}].

\bibitem{Afik:2023mhj}
Y.~Afik, B.~D\"obrich, J.~Jerhot, Y.~Soreq and K.~Tobioka, \emph{{Probing
  Long-lived Axions at the KOTO Experiment}},
  \href{https://arxiv.org/abs/2303.01521}{{\ttfamily 2303.01521}}.

\bibitem{Shifman:1979if}
M.A.~Shifman, A.I.~Vainshtein and V.I.~Zakharov, \emph{{Can Confinement Ensure
  Natural CP Invariance of Strong Interactions?}},
  \href{https://doi.org/10.1016/0550-3213(80)90209-6}{\emph{Nucl. Phys. B}
  {\bfseries 166} (1980) 493}.

\bibitem{Zhitnitsky:1980tq}
A.R.~Zhitnitsky, \emph{{On Possible Suppression of the Axion Hadron
  Interactions. (In Russian)}}, {\emph{Sov. J. Nucl. Phys.} {\bfseries 31}
  (1980) 260}.

\bibitem{Dine:1981rt}
M.~Dine, W.~Fischler and M.~Srednicki, \emph{{A Simple Solution to the Strong
  CP Problem with a Harmless Axion}},
  \href{https://doi.org/10.1016/0370-2693(81)90590-6}{\emph{Phys. Lett. B}
  {\bfseries 104} (1981) 199}.

\bibitem{Srednicki:1985xd}
M.~Srednicki, \emph{{Axion Couplings to Matter. 1. CP Conserving Parts}},
  \href{https://doi.org/10.1016/0550-3213(85)90054-9}{\emph{Nucl. Phys. B}
  {\bfseries 260} (1985) 689}.

\bibitem{Sun:2020iim}
J.~Sun and X.-G.~He, \emph{{DFSZ axion couplings revisited}},
  \href{https://doi.org/10.1016/j.physletb.2020.135881}{\emph{Phys. Lett. B}
  {\bfseries 811} (2020) 135881}
  [\href{https://arxiv.org/abs/2006.16931}{{\ttfamily 2006.16931}}].

\bibitem{Chala:2020wvs}
M.~Chala, G.~Guedes, M.~Ramos and J.~Santiago, \emph{{Running in the ALPs}},
  \href{https://doi.org/10.1140/epjc/s10052-021-08968-2}{\emph{Eur. Phys. J. C}
  {\bfseries 81} (2021) 181}
  [\href{https://arxiv.org/abs/2012.09017}{{\ttfamily 2012.09017}}].

\bibitem{Gavela:2019wzg}
M.B.~Gavela, R.~Houtz, P.~Quilez, R.~Del~Rey and O.~Sumensari, \emph{{Flavor
  constraints on electroweak ALP couplings}},
  \href{https://doi.org/10.1140/epjc/s10052-019-6889-y}{\emph{Eur. Phys. J. C}
  {\bfseries 79} (2019) 369}
  [\href{https://arxiv.org/abs/1901.02031}{{\ttfamily 1901.02031}}].

\bibitem{DiLuzio:2020oah}
L.~Di~Luzio, R.~Gr\"ober and P.~Paradisi, \emph{{Hunting for $CP$-violating
  axionlike particle interactions}},
  \href{https://doi.org/10.1103/PhysRevD.104.095027}{\emph{Phys. Rev. D}
  {\bfseries 104} (2021) 095027}
  [\href{https://arxiv.org/abs/2010.13760}{{\ttfamily 2010.13760}}].

\bibitem{Giraldo:2020hwl}
Y.~Giraldo, R.~Martinez, E.~Rojas and J.C.~Salazar, \emph{{Flavored axions and
  the flavor problem}},
  \href{https://doi.org/10.1140/epjc/s10052-022-11073-7}{\emph{Eur. Phys. J. C}
  {\bfseries 82} (2022) 1131}
  [\href{https://arxiv.org/abs/2007.05653}{{\ttfamily 2007.05653}}].

\bibitem{Song:2023lxf}
H.~Song, H.~Sun and J.-H.~Yu, \emph{{Effective Field Theories of Axion, ALP and
  Dark Photon}},  \href{https://arxiv.org/abs/2305.16770}{{\ttfamily
  2305.16770}}.

\bibitem{Chang:2000ii}
D.~Chang, W.-F.~Chang, C.-H.~Chou and W.-Y.~Keung, \emph{{Large two loop
  contributions to g-2 from a generic pseudoscalar boson}},
  \href{https://doi.org/10.1103/PhysRevD.63.091301}{\emph{Phys. Rev. D}
  {\bfseries 63} (2001) 091301}
  [\href{https://arxiv.org/abs/hep-ph/0009292}{{\ttfamily hep-ph/0009292}}].

\bibitem{Buen-Abad:2021fwq}
M.A.~Buen-Abad, J.~Fan, M.~Reece and C.~Sun, \emph{{Challenges for an axion
  explanation of the muon $g - 2$ measurement}},
  \href{https://doi.org/10.1007/JHEP09(2021)101}{\emph{JHEP} {\bfseries 09}
  (2021) 101} [\href{https://arxiv.org/abs/2104.03267}{{\ttfamily
  2104.03267}}].

\bibitem{Marciano:2016yhf}
W.J.~Marciano, A.~Masiero, P.~Paradisi and M.~Passera, \emph{{Contributions of
  axionlike particles to lepton dipole moments}},
  \href{https://doi.org/10.1103/PhysRevD.94.115033}{\emph{Phys. Rev. D}
  {\bfseries 94} (2016) 115033}
  [\href{https://arxiv.org/abs/1607.01022}{{\ttfamily 1607.01022}}].

\bibitem{Cornella:2019uxs}
C.~Cornella, P.~Paradisi and O.~Sumensari, \emph{{Hunting for ALPs with Lepton
  Flavor Violation}},
  \href{https://doi.org/10.1007/JHEP01(2020)158}{\emph{JHEP} {\bfseries 01}
  (2020) 158} [\href{https://arxiv.org/abs/1911.06279}{{\ttfamily
  1911.06279}}].

\bibitem{Bauer:2019gfk}
M.~Bauer, M.~Neubert, S.~Renner, M.~Schnubel and A.~Thamm, \emph{{Axionlike
  Particles, Lepton-Flavor Violation, and a New Explanation of $a_\mu$ and
  $a_e$}}, \href{https://doi.org/10.1103/PhysRevLett.124.211803}{\emph{Phys.
  Rev. Lett.} {\bfseries 124} (2020) 211803}
  [\href{https://arxiv.org/abs/1908.00008}{{\ttfamily 1908.00008}}].

\bibitem{Liu:2022tqn}
J.~Liu, X.~Ma, L.-T.~Wang and X.-P.~Wang, \emph{{The ALP explanation to muon
  $(g-2)$ and its test at future Tera-$Z$ and Higgs factories}},
  \href{https://arxiv.org/abs/2210.09335}{{\ttfamily 2210.09335}}.

\bibitem{Essig:2010gu}
R.~Essig, R.~Harnik, J.~Kaplan and N.~Toro, \emph{{Discovering New Light States
  at Neutrino Experiments}},
  \href{https://doi.org/10.1103/PhysRevD.82.113008}{\emph{Phys. Rev. D}
  {\bfseries 82} (2010) 113008}
  [\href{https://arxiv.org/abs/1008.0636}{{\ttfamily 1008.0636}}].

\bibitem{Liu:2017htz}
Y.-S.~Liu and G.A.~Miller, \emph{{Validity of the Weizs\"acker-Williams
  approximation and the analysis of beam dump experiments: Production of an
  axion, a dark photon, or a new axial-vector boson}},
  \href{https://doi.org/10.1103/PhysRevD.96.016004}{\emph{Phys. Rev. D}
  {\bfseries 96} (2017) 016004}
  [\href{https://arxiv.org/abs/1705.01633}{{\ttfamily 1705.01633}}].

\bibitem{DiLuzio:2020wdo}
L.~Di~Luzio, M.~Giannotti, E.~Nardi and L.~Visinelli, \emph{{The landscape of
  QCD axion models}},
  \href{https://doi.org/10.1016/j.physrep.2020.06.002}{\emph{Phys. Rept.}
  {\bfseries 870} (2020) 1} [\href{https://arxiv.org/abs/2003.01100}{{\ttfamily
  2003.01100}}].

\bibitem{Georgi:1986df}
H.~Georgi, D.B.~Kaplan and L.~Randall, \emph{{Manifesting the Invisible Axion
  at Low-energies}},
  \href{https://doi.org/10.1016/0370-2693(86)90688-X}{\emph{Phys. Lett. B}
  {\bfseries 169} (1986) 73}.

\bibitem{Chang:1993gm}
S.~Chang and K.~Choi, \emph{{Hadronic axion window and the big bang
  nucleosynthesis}},
  \href{https://doi.org/10.1016/0370-2693(93)90656-3}{\emph{Phys. Lett. B}
  {\bfseries 316} (1993) 51}
  [\href{https://arxiv.org/abs/hep-ph/9306216}{{\ttfamily hep-ph/9306216}}].

\bibitem{Grossman:1994jb}
Y.~Grossman, \emph{{Phenomenology of models with more than two Higgs
  doublets}}, \href{https://doi.org/10.1016/0550-3213(94)90316-6}{\emph{Nucl.
  Phys. B} {\bfseries 426} (1994) 355}
  [\href{https://arxiv.org/abs/hep-ph/9401311}{{\ttfamily hep-ph/9401311}}].

\bibitem{Akeroyd:1994ga}
A.G.~Akeroyd and W.J.~Stirling, \emph{{Light charged Higgs scalars at
  high-energy $e^+ e^-$ colliders}},
  \href{https://doi.org/10.1016/0550-3213(95)00173-P}{\emph{Nucl. Phys. B}
  {\bfseries 447} (1995) 3}.

\bibitem{Akeroyd:1996he}
A.G.~Akeroyd, \emph{{Nonminimal neutral Higgs bosons at LEP-2}},
  \href{https://doi.org/10.1016/0370-2693(96)00330-9}{\emph{Phys. Lett. B}
  {\bfseries 377} (1996) 95}
  [\href{https://arxiv.org/abs/hep-ph/9603445}{{\ttfamily hep-ph/9603445}}].

\bibitem{Aoki:2009ha}
M.~Aoki, S.~Kanemura, K.~Tsumura and K.~Yagyu, \emph{{Models of Yukawa
  interaction in the two Higgs doublet model, and their collider
  phenomenology}},
  \href{https://doi.org/10.1103/PhysRevD.80.015017}{\emph{Phys. Rev. D}
  {\bfseries 80} (2009) 015017}
  [\href{https://arxiv.org/abs/0902.4665}{{\ttfamily 0902.4665}}].

\bibitem{Branco:2011iw}
G.C.~Branco, P.M.~Ferreira, L.~Lavoura, M.N.~Rebelo, M.~Sher and J.P.~Silva,
  \emph{{Theory and phenomenology of two-Higgs-doublet models}},
  \href{https://doi.org/10.1016/j.physrep.2012.02.002}{\emph{Phys. Rept.}
  {\bfseries 516} (2012) 1} [\href{https://arxiv.org/abs/1106.0034}{{\ttfamily
  1106.0034}}].

\bibitem{Bhattacharyya:2015nca}
G.~Bhattacharyya and D.~Das, \emph{{Scalar sector of two-Higgs-doublet models:
  A minireview}},
  \href{https://doi.org/10.1007/s12043-016-1252-4}{\emph{Pramana} {\bfseries
  87} (2016) 40} [\href{https://arxiv.org/abs/1507.06424}{{\ttfamily
  1507.06424}}].

\bibitem{Liu:2018xkx}
J.~Liu, C.E.M.~Wagner and X.-P.~Wang, \emph{{A light complex scalar for the
  electron and muon anomalous magnetic moments}},
  \href{https://doi.org/10.1007/JHEP03(2019)008}{\emph{JHEP} {\bfseries 03}
  (2019) 008} [\href{https://arxiv.org/abs/1810.11028}{{\ttfamily
  1810.11028}}].

\bibitem{Spira:1995rr}
M.~Spira, A.~Djouadi, D.~Graudenz and P.M.~Zerwas, \emph{{Higgs boson
  production at the LHC}},
  \href{https://doi.org/10.1016/0550-3213(95)00379-7}{\emph{Nucl. Phys. B}
  {\bfseries 453} (1995) 17}
  [\href{https://arxiv.org/abs/hep-ph/9504378}{{\ttfamily hep-ph/9504378}}].

\bibitem{Bjorkeroth:2019jtx}
F.~Bj\"orkeroth, L.~Di~Luzio, F.~Mescia, E.~Nardi, P.~Panci and R.~Ziegler,
  \emph{{Axion-electron decoupling in nucleophobic axion models}},
  \href{https://doi.org/10.1103/PhysRevD.101.035027}{\emph{Phys. Rev. D}
  {\bfseries 101} (2020) 035027}
  [\href{https://arxiv.org/abs/1907.06575}{{\ttfamily 1907.06575}}].

\bibitem{DiLuzio:2016sur}
L.~Di~Luzio, J.F.~Kamenik and M.~Nardecchia, \emph{{Implications of
  perturbative unitarity for scalar di-boson resonance searches at LHC}},
  \href{https://doi.org/10.1140/epjc/s10052-017-4594-2}{\emph{Eur. Phys. J. C}
  {\bfseries 77} (2017) 30} [\href{https://arxiv.org/abs/1604.05746}{{\ttfamily
  1604.05746}}].

\bibitem{DiLuzio:2017chi}
L.~Di~Luzio and M.~Nardecchia, \emph{{What is the scale of new physics behind
  the $B$-flavour anomalies?}},
  \href{https://doi.org/10.1140/epjc/s10052-017-5118-9}{\emph{Eur. Phys. J. C}
  {\bfseries 77} (2017) 536}
  [\href{https://arxiv.org/abs/1706.01868}{{\ttfamily 1706.01868}}].

\bibitem{Alloul:2013bka}
A.~Alloul, N.D.~Christensen, C.~Degrande, C.~Duhr and B.~Fuks, \emph{{FeynRules
  2.0 - A complete toolbox for tree-level phenomenology}},
  \href{https://doi.org/10.1016/j.cpc.2014.04.012}{\emph{Comput. Phys. Commun.}
  {\bfseries 185} (2014) 2250}
  [\href{https://arxiv.org/abs/1310.1921}{{\ttfamily 1310.1921}}].

\bibitem{Mertig:1990an}
R.~Mertig, M.~Bohm and A.~Denner, \emph{{FEYN CALC: Computer algebraic
  calculation of Feynman amplitudes}},
  \href{https://doi.org/10.1016/0010-4655(91)90130-D}{\emph{Comput. Phys.
  Commun.} {\bfseries 64} (1991) 345}.

\bibitem{Shtabovenko:2016sxi}
V.~Shtabovenko, R.~Mertig and F.~Orellana, \emph{{New Developments in FeynCalc
  9.0}}, \href{https://doi.org/10.1016/j.cpc.2016.06.008}{\emph{Comput. Phys.
  Commun.} {\bfseries 207} (2016) 432}
  [\href{https://arxiv.org/abs/1601.01167}{{\ttfamily 1601.01167}}].

\bibitem{Shtabovenko:2020gxv}
V.~Shtabovenko, R.~Mertig and F.~Orellana, \emph{{FeynCalc 9.3: New features
  and improvements}},
  \href{https://doi.org/10.1016/j.cpc.2020.107478}{\emph{Comput. Phys. Commun.}
  {\bfseries 256} (2020) 107478}
  [\href{https://arxiv.org/abs/2001.04407}{{\ttfamily 2001.04407}}].

\bibitem{Hahn:2000kx}
T.~Hahn, \emph{{Generating Feynman diagrams and amplitudes with FeynArts 3}},
  \href{https://doi.org/10.1016/S0010-4655(01)00290-9}{\emph{Comput. Phys.
  Commun.} {\bfseries 140} (2001) 418}
  [\href{https://arxiv.org/abs/hep-ph/0012260}{{\ttfamily hep-ph/0012260}}].

\bibitem{Abashian:1980pb}
A.~Abashian, J.D.~Bjorken, L.W.~Mo, W.R.~Nelson and Y.-S.~Tsai, \emph{{Proposal
  to Search for Low Mass, Metastable, Neutral Particles at {SLAC}}}, .

\bibitem{Ishikawa:2021qna}
A.~Ishikawa, Y.~Sakaki and Y.~Takubo, \emph{{Search for axion-like particles
  with electron and positron beams at the KEK linac}},
  \href{https://doi.org/10.1093/ptep/ptac129}{\emph{PTEP} {\bfseries 2022}
  (2022) 113B05} [\href{https://arxiv.org/abs/2107.06431}{{\ttfamily
  2107.06431}}].

\bibitem{Riordan:1987aw}
E.M.~Riordan et~al., \emph{{A Search for Short Lived Axions in an Electron Beam
  Dump Experiment}},
  \href{https://doi.org/10.1103/PhysRevLett.59.755}{\emph{Phys. Rev. Lett.}
  {\bfseries 59} (1987) 755}.

\bibitem{Bross:1989mp}
A.~Bross, M.~Crisler, S.H.~Pordes, J.~Volk, S.~Errede and J.~Wrbanek, \emph{{A
  Search for Shortlived Particles Produced in an Electron Beam Dump}},
  \href{https://doi.org/10.1103/PhysRevLett.67.2942}{\emph{Phys. Rev. Lett.}
  {\bfseries 67} (1991) 2942}.

\bibitem{Davier:1989wz}
M.~Davier and H.~Nguyen~Ngoc, \emph{{An Unambiguous Search for a Light Higgs
  Boson}}, \href{https://doi.org/10.1016/0370-2693(89)90174-3}{\emph{Phys.
  Lett. B} {\bfseries 229} (1989) 150}.

\bibitem{Andreas:2012mt}
S.~Andreas, C.~Niebuhr and A.~Ringwald, \emph{{New Limits on Hidden Photons
  from Past Electron Beam Dumps}},
  \href{https://doi.org/10.1103/PhysRevD.86.095019}{\emph{Phys. Rev. D}
  {\bfseries 86} (2012) 095019}
  [\href{https://arxiv.org/abs/1209.6083}{{\ttfamily 1209.6083}}].

\bibitem{Tsai:1966js}
Y.-S.~Tsai and V.~Whitis, \emph{{Thick-Target Bremsstrahlung and Target
  Considerations for Secondary-Particle Production by Electrons}},
  \href{https://doi.org/10.1103/PhysRev.149.1248}{\emph{Phys. Rev.} {\bfseries
  149} (1966) 1248}.

\bibitem{Kim:1973he}
K.J.~Kim and Y.-S.~Tsai, \emph{{Improved Weizs\"acker-Williams Method and Its
  Application to Lepton and W-Boson Pair Production}},
  \href{https://doi.org/10.1103/PhysRevD.8.3109}{\emph{Phys. Rev. D} {\bfseries
  8} (1973) 3109}.

\bibitem{Tsai:1973py}
Y.-S.~Tsai, \emph{{Pair Production and Bremsstrahlung of Charged Leptons}},
  \href{https://doi.org/10.1103/RevModPhys.46.815}{\emph{Rev. Mod. Phys.}
  {\bfseries 46} (1974) 815}.

\bibitem{Tsai:1986tx}
Y.-S.~Tsai, \emph{{Axion bremsstrahlung by an electron beam}},
  \href{https://doi.org/10.1103/PhysRevD.34.1326}{\emph{Phys. Rev. D}
  {\bfseries 34} (1986) 1326}.

\bibitem{Bauer:2018onh}
M.~Bauer, P.~Foldenauer and J.~Jaeckel, \emph{{Hunting All the Hidden
  Photons}}, \href{https://doi.org/10.1007/JHEP07(2018)094}{\emph{JHEP}
  {\bfseries 07} (2018) 094}
  [\href{https://arxiv.org/abs/1803.05466}{{\ttfamily 1803.05466}}].

\bibitem{Gninenko:2017yus}
S.N.~Gninenko, D.V.~Kirpichnikov, M.M.~Kirsanov and N.V.~Krasnikov, \emph{{The
  exact tree-level calculation of the dark photon production in high-energy
  electron scattering at the CERN SPS}},
  \href{https://doi.org/10.1016/j.physletb.2018.05.010}{\emph{Phys. Lett. B}
  {\bfseries 782} (2018) 406}
  [\href{https://arxiv.org/abs/1712.05706}{{\ttfamily 1712.05706}}].

\bibitem{Bjorken:2009mm}
J.D.~Bjorken, R.~Essig, P.~Schuster and N.~Toro, \emph{{New Fixed-Target
  Experiments to Search for Dark Gauge Forces}},
  \href{https://doi.org/10.1103/PhysRevD.80.075018}{\emph{Phys. Rev. D}
  {\bfseries 80} (2009) 075018}
  [\href{https://arxiv.org/abs/0906.0580}{{\ttfamily 0906.0580}}].

\bibitem{Schiff:1951zza}
L.I.~Schiff, \emph{{Energy-Angle Distribution of Thin Target Bremsstrahlung}},
  \href{https://doi.org/10.1103/PhysRev.83.252}{\emph{Phys. Rev.} {\bfseries
  83} (1951) 252}.

\bibitem{Liu:2016mqv}
Y.-S.~Liu, D.~McKeen and G.A.~Miller, \emph{{Validity of the
  Weizs\"acker-Williams approximation and the analysis of beam dump
  experiments: Production of a new scalar boson}},
  \href{https://doi.org/10.1103/PhysRevD.95.036010}{\emph{Phys. Rev. D}
  {\bfseries 95} (2017) 036010}
  [\href{https://arxiv.org/abs/1609.06781}{{\ttfamily 1609.06781}}].

\bibitem{EDELWEISS:2018tde}
{\scshape EDELWEISS} collaboration, \emph{{Searches for electron interactions
  induced by new physics in the EDELWEISS-III Germanium bolometers}},
  \href{https://doi.org/10.1103/PhysRevD.98.082004}{\emph{Phys. Rev. D}
  {\bfseries 98} (2018) 082004}
  [\href{https://arxiv.org/abs/1808.02340}{{\ttfamily 1808.02340}}].

\end{thebibliography}\endgroup

\end{document}